\documentclass[lettersize,journal]{IEEEtran}
\usepackage{mathtools}
\usepackage{setspace}
\usepackage{amsmath}
\usepackage{amsmath,amsfonts}
\usepackage{algorithmic}
\usepackage{algorithm}
\usepackage{array}
\usepackage[caption=false,font=normalsize,labelfont=sf,textfont=sf]{subfig}
\usepackage{textcomp}
\usepackage{stfloats}
\usepackage{url}
\usepackage{verbatim}
\usepackage{graphicx}
\usepackage{amsthm,amsmath,amssymb}
\usepackage{mathrsfs}
\usepackage{booktabs}
\usepackage{amssymb}
\usepackage[noadjust]{cite}

\usepackage{changepage}
\begin{document}

\title{Wireless Powered Metaverse: Joint Task Scheduling and Trajectory Design for Multi-Devices and Multi-UAVs}

\author{Xiaojie Wang, Jiameng Li, Zhaolong Ning, Qingyang Song,\\ Lei Guo, Abbas Jamalipour, \textit{Fellow, IEEE}
	
	\thanks{This work was supported by the Natural Science Foundation ofChina under Grants 61971084, 62025105, 62001073, 62221005 and
62272075, by the National Natural Science Foundation of Chongqing
under Grants cstc2021ycjhbgzxm0039, CSTB2022BSXM-JCX0109, and
CSTB2022BSXMJCX0110, by the Science and Technology Research Program for Chongqing Municipal Education Commission KJZDM202200601, by the Support Program for Overseas Students to Return to China for Entrepreneurship and Innovation under Grants cx2021003 and cx2021053, and by the Youth Innovation Group Support Program of ICE Discipline of CQUPT (SCIE-QN-2022-03). \textit{(Corresponding author: Zhaolong Ning.)}\\
		\indent X. Wang, J. Li, Z. Ning , Q. Song, and L. Guo are with the School of Communications and Information Engineering, Chongqing University of Posts and Telecommunications, Chongqing, China. Email: xiaojie.kara.wang@ieee.org, s210101075@stu.cqupt.edu.cn, z.ning@ieee.org, songqy@cqupt.edu.cn, guolei@cqupt.edu.cn.\\ \indent A. Jamalipour is with the School of Electrical and lnformation Engineering, The University of Sydney, Sydney, Australia. Email: a.jamalipour@ieee.org.
	}
}


\maketitle

\begin{abstract}
	
To support the running of human-centric metaverse applications on mobile devices, Unmanned Aerial Vehicle (UAV)-assisted Wireless Powered Mobile Edge Computing (WPMEC) is promising to compensate for limited computational capabilities and energy supplies of mobile devices. 
The high-speed computational processing demands and significant energy consumption of metaverse applications require joint resource scheduling of multiple devices and UAVs, but existing WPMEC solutions address either device or UAV scheduling due to the complexity of combinatorial optimization. 
To solve the above challenge, we propose a two-stage alternating optimization algorithm based on multi-task Deep Reinforcement Learning (DRL) to jointly allocate charging time, schedule computation tasks, and optimize trajectory of UAVs and mobile devices in a wireless powered metaverse scenario.
First, considering energy constraints of both UAVs and mobile devices, we formulate an optimization problem to maximize the computation efficiency of the system. 
Second, we propose a heuristic algorithm to efficiently perform time allocation and charging scheduling for mobile devices. 
Following this, we design a multi-task DRL scheme to make charging scheduling and trajectory design decisions for UAVs. 
Finally, theoretical analysis and performance results demonstrate that our algorithm exhibits significant advantages over representative methods in terms of convergence speed and average computation efficiency.

\end{abstract}

\section{Introduction}
\IEEEPARstart{T}{he} metaverse, referred as the successor of mobile Internet~\cite{ref19}, is gaining popularity as a concept.
It is the materialized version of the Internet, which includes a human-centric, immersive, and interoperable virtual ecosystem that can be navigated by virtual characters controlled by users~\cite{ref18}. 
In the metaverse, users can interact with the virtual world through body movements or sound, and experience comprehensive human-centric virtual services. 
For example, users can explore the virtual world in the metaverse with the help of technologies such as virtual reality, augmented reality and mixed reality.
However, the key characteristic of these technologies is timely rendering of images to quickly generate perceptual images based on users' thoughts, requiring devices to have high computing capability and sufficient energy.\\
\indent With the development of metaverse-related technologies and the rising demand of users to perceive virtual worlds dynamically in real time, traditional wired devices cannot realize ubiquitous network access to the metaverse. 
Instead, running programs developed for the metaverse at mobile devices held by users can create a human-centric and dynamic virtual world perceiving users' needs in real time~\cite{ref19}. 
Each user creates his customized avatar and interacts with other avatars and digital objects for a more realistic experience in the metaverse anytime and anywhere, inevitably posing significant challenges to the computing capability and battery capacity of each terminal.

\indent Currently, due to production costs and technological limitations, computing capability and battery life of mobile devices are always constrained \cite{8967118}. 
In recent years, Mobile Edge Computing (MEC) \cite{9210202} has been regarded as a key solution to support real-time rendering and become one of the feasible solutions to provide real-time virtual reality videos through wireless networks. 
It works as a strong supplement to the computing capability of mobile devices by offloading computation tasks to nearby MEC servers \cite{8887204}. 
Additionally, with the development of Wireless Power Transfer (WPT) technology, the energy constraint of mobile devices has been alleviated.
Therefore, with the integration of WPT and MEC, Wireless Powered Mobile Edge Computing (WPMEC) lifts the burden of battery life while enhancing the computing capability for sustaining the metaverse~\cite{ref21}. 
With the aid of the WPMEC technology, running human-centric programs developed for the metaverse at mobile devices becomes smooth and easy, and we refer to this paradigm as wireless powered metaverse.

\indent Due to their flexibility and controllability, Unmanned Aerial Vehicles (UAVs) with controllable trajectories are often used as airborne edge servers that can dynamically provide computing services to mobile devices~\cite{ref23,10239498}. Although some studies have discussed UAV-assisted WPMEC (\cite{ref35,ref36,ref37,ref6}), they cannot be directly applied to support human-centric metaverse applications due to the following reasons:
\begin{itemize}{}{}
	\item{First, there is an urgent requirement to provide timely energy supply and efficient allocation of network resources, especially when large-scale dynamic coexistence of mobile devices with metaverse access requirements.
However, existing research usually only considers charging and task scheduling issues of mobile devices, while ignoring or simplifying the similar issues of UAVs. 
 Therefore, it is necessary to design a novel trajectory optimization and task scheduling scheme for both UAVs and mobile devices.}
\item{Second, human-centric metaverse applications require significant computational resources and high computing speeds.
However, the half-duplex characteristic of devices and the charging requirement of UAVs can cause instability in the amount of computational resources in a WPMEC network. 
Therefore, it is worth studying how to allocate the charging time of UAVs and mobile devices reasonably, to maximize the computation efficiency of the system.
}
\item{Finally, to address the optimization problem of UAV-assisted WPMEC systems, existing research mostly adopts classical convex optimization methods or heuristic algorithms. 
	However, in human-centric metaverse scenarios, multiple coupled scheduling parameters and rapidly changing network states make the traditional optimization algorithms inefficient.
	Therefore, to enable smooth operations of metaverse programs at mobile devices, efficient learning algorithms are needed for rational resource scheduling.
}
\end{itemize}

\indent To address the above challenges, we propose a \underline{mu}lti-task deep \underline{r}einforcement le\underline{a}rning based two-stage a\underline{l}ternating optimization algorithm, named MURAL, which can be applied to a multi-UAV-assisted WPMEC network with human-centric metaverse devices.
With the optimization goal of maximizing the computation efficiency of the system, this algorithm allows for joint charging time allocation, computation offloading, and UAV trajectory design.
To the best of our knowledge, this is \textbf{\textit{the first work that considers joint charging time allocation, computation task scheduling and trajectory optimization for both UAVs and mobile devices. }}
A summary of our contribution can be found below:
\begin{itemize}{}{}
	\item{We construct a system model based on WPMEC, which allows simultaneous charging of both  UAVs and mobile devices, providing support for the operation of programs developed for the metaverse. 
		To achieve efficient utilization of system resources, we propose a novel time allocation model for UAVs and mobile devices by considering their half-duplex characteristics.}
	\item{Given energy constraints of mobile devices and UAVs, we formulate an optimization problem for maximizing the long-term system computation efficiency. 
		To solve this problem, we first decompose it into two subproblems: time allocation and charging scheduling for mobile devices, and charging scheduling and trajectory design for UAVs. 
		We propose a two-stage alternating optimization algorithm based on multi-task Deep Reinforcement Learning (DRL) to solve these subproblems.
}
	\item{For the first subproblem, we design a heuristic and effective algorithm and theoretically prove the existence of an optimal solution.
	For the second subproblem, we propose a learning algorithm based on multi-task DRL to learn strategies for different UAVs, and theoretically show that the complexity of this approach is polynomial time.}
	\item{We conduct performance evaluation of the designed algorithm based on the Manhattan city map and compare it against several representative algorithms. 
		The evaluation results verify the effectiveness of our approach in terms of convergence speed, average computation efficiency, average numbers of computation bits and average energy consumption. 	
}
\end{itemize}

The rest of this paper is organized as follows: In Section \ref{2}, we illustrate the related work; we construct the system model and formulate the optimization problem in Section \ref{3}; In Section \ref{4}, we design the two-stage alternating optimization algorithm based on multi-task DRL; The performance evaluation is conducted in Section \ref{5}; Finally, we conclude this work in Section \ref{6}.

\section{Related Work}\label{2}
In this section, we review the related work  including UAV-assisted WPMEC and multi-task DRL.

\subsection{UAV-Assisted WPMEC}
Integrating UAV communication technology with WPMEC systems can mitigate the drawbacks of fixed energy supply and user mobility~\cite{ref35,ref36,ref37,ref6}. 
Considering the randomness of energy and computation tasks in the spatial domain, authors in~\cite{ref35} investigated resource allocation of UAV-assisted WPMEC networks to minimize UAV energy consumption subject to stable data and energy queues. 
In~\cite{ref36}, authors applied non-orthogonal multiple access and hybrid beamforming techniques to maximize the computation efficiency of UAV-assisted WPMEC systems.
While the aforementioned studies focused on optimizing UAV trajectories and resource allocation for UAV-assisted WPMEC systems, none of them considered UAV charging scheduling.

\indent Recently, a few researchers have utilized laser charging to alleviate energy consumption in UAV-assisted WPMEC networks.
Authors in~\cite{ref37} designed a UAV-assisted user offloading scheme and employed laser charging for UAVs. 
They formulated a resource allocation and trajectory optimization problem, considering the UAV's service time and task completion time.
In addition, authors in~\cite{ref6} assumed UAVs as information and energy relay stations to assist with task offloading and energy transfer.
Although the above studies considered the UAV charging issue, they only investigated a single UAV-assisted WPMEC network and used traditional optimization algorithms that are not suitable for dynamic environments.
\subsection{Multi-Task DRL}
Due to the limitation of traditional DRL that only learns the control policies of a single task, and cannot solve multiple related problems simultaneously or sequentially with reasonable time consumption and satisfactory performance, researchers (~\cite{ref39,ref9,ref40}) have begun to explore multi-task DRL that allows agents to learn multiple sequential decision tasks at once and performs well on different tasks~\cite{ref40}.
The multi-task DRL approach has the natural advantage of solving multiple tasks with the same structure and interconnections in parallel, and can increase the generalization capability of the model with shared knowledge learned by intelligences.

\begin{figure*}[!t]
	\centering
	\includegraphics[width=5.8in]{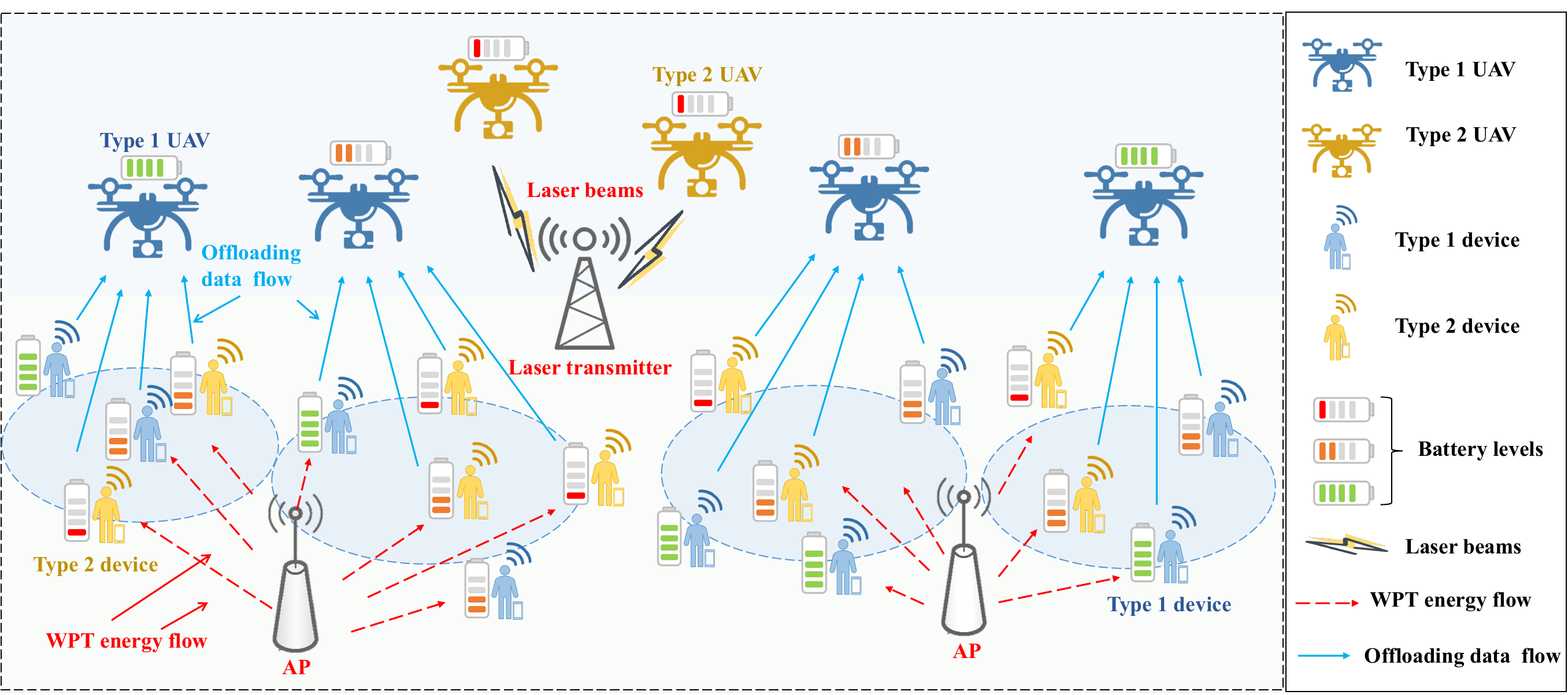}
	\caption{The illustrative system model.}
	\label{11}
\end{figure*}

\indent Distral~\cite{ref9} is a classic multi-task DRL approach that combines knowledge distillation and transfer learning, purifying strategies learned on each task to obtain a shared strategy, and then using this shared strategy to guide the strategy on each specific task for better learning.
$ \pi_i=(S,A,p_i(s'|s,a),\gamma,R_i(a,s)) $ denotes the policy for corresponding task $ i $, and $ \pi_0 $ represents the shared policy obtained after refinement of tasks $\{ \pi_i \}_{i=1}^n$. 
Therefore, the objective to be maximized is defined as follows:
\begin{align*}\label{jpi}
	&J(\pi_0,\{\pi_i\}_{i=1}^n)\\&=\sum_{i} \mathbb{E}_{\pi_i}\sum_{t\geq0}\Big(\gamma^tR_i\left(a_t\middle| s_t\right)-c_{KL}\gamma^t\log{\pi_0\left(a_t\middle| s_t\right)}
	\\&-(c_{Ent}-c_{KL})\gamma^t\log{\pi_i\left(a_t\middle| s_t\right)} \Big), \tag{1}
\end{align*}
where variables $ c_{KL}$ and  $ c_{Ent}  $ are scalar factors determining the strengths of Kullback-Leibler (KL) and entropy regularization, respectively.

\indent The multi-task DRL approach has a wide range of applications, such as the modelling of partial differential equations~\cite{2018Belletti} and the embedding of knowledge graphs~\cite{2023Zhang}.
Authors in~\cite{ref31} proposed a parallel task scheduling algorithm based on multi-task DRL in the Internet of things, taking the correlation among tasks  into account.
Authors in~\cite{ref28} used a multi-task DRL algorithm based on graph convolutional networks to achieve network slicing and routing.
In addition, authors in~\cite{ref29} proposed an intelligent resource allocation framework based on multi-task DRL to implement distributed computing.
Although multi-task DRL has great potential in solving resource allocation problems of  wireless networks, there has been little research on joint task scheduling and trajectory design based on multi-task DRL in UAV-assisted WPMEC networks.

\indent In this paper, we propose MURAL, a joint scheduling algorithm suitable for multi-UAV assisted WPMEC networks, which is designed for mobile devices supporting for the running of human-centric programs developed for the metaverse and aims to maximize the computation efficiency of the system.
To the best of our knowledge, this is \textit{the first work that considers joint charging time allocation, computation task scheduling  and trajectory design for both UAVs and mobile devices.}
\begin{table}[!t]
	\centering
	\caption{{Main Notations}}
	\label{canshu1}
	\begin{tabular}{ll}
		\toprule[0.8pt]
		{Notation}    \quad\quad\quad                                                                      & {Description}  \\ 
		\toprule[0.8pt]
		$ q_m^t$                                                                  & The coordinate of mobile device $m$ in  \\ & time slot $ t $;             \\
		$ q_u^t$                                                                  & The coordinate of UAV $u$ in time  slot $ t $;             \\
		$ h_{mu}^t $&The channel gain between UAV $u$ and \\ & mobile device $m$ in time slot $ t $; \\
		$ \alpha_m^t $&The scheduling variable of device $ m $ in \\ & time slot $ t $;  \\
		$ \beta_u^t $&The scheduling variable of UAV $ u $ in \\ & time slot $ t $;  \\
		$ \tau^t $&	The time allocation variable in time slot $ t $;\\
		$\varepsilon$& The energy conversion efficiency of mobile \\ & devices;\\
		$\epsilon$ &The energy conversion efficiency for UAVs;\\
		$\mathcal{P}_{1}$ &The transmission power of the laser;\\
		$P_m$ &The transmission power of device $m$;\\
		$ T $&The duration of a time slot;\\ 
		$C_1$ &The number of CPU cycles required to \\ & compute each data bit;\\
		$f_m$& The CPU frequency of device $m$;\\
		$f_u$& The CPU frequency of UAV $u$;\\
		$k_m$ &The coefficient related to the computation \\ & efficiency of device $m$;\\
		$k_u$& The coefficient related to the computation\\ & efficiency of UAV $u$;\\
		${E}_{m}^{t,r}  $& The residual battery energy of mobile \\ & device $m$ in time slot $ t $;\\
		${\mathbb{E}}_{u}^{t,r}  $& The residual battery energy of UAV $u$ in \\ & time slot $ t $;\\
		$ s_u^t$& The state of UAV $ u $ in time slot $ t $;\\ 
		$ s^t $&The state of all UAVs in time slot $ t $;\\
		$ a_u^t$& The action of UAV $ u $ in time slot $ t $;\\ 
		$ a^t$& The joint action of UAVs in time slot $ t $;\\ 
		$ R_u\left(a_u^t\middle| s^t\right) $&The immediate reward received by UAV $ u $ \\ & after taking action $ a_u^t $ at state $  s^t$;\\
		$ \pi_0 $& The shared policy in the \\ & designed learning algorithm;\\
		$ \pi_u $&The task-specific policy of UAV $ u $ in the \\ & designed learning algorithm;\\
		$ \theta_0 $, $ \{\theta_u\}_{u \in \mathbb{U}}$&Optimization parameters of the shared\\ & policy  network and task-specific policy\\ & networks.\\
		\toprule[0.8pt]  
	\end{tabular}
\end{table}

\section{System Model and Problem Formulation}\label{3}
\indent In this section, we present the constructed system model and formulate the long-term computation efficiency maximization problem.
The main notations can be found in Table \ref{canshu1}.

\subsection{System Model}
We consider a UAV-assisted WPMEC system, which can provide support for running programs developed for the metaverse. 
As shown in Fig. \ref{11}, we consider a three-dimensional region containing Access Points (APs), laser emitters, $ U $ UAVs, denoted by $\mathbb{U} = \{1, ..., u, ..., U\}$, and $ M $ mobile devices, represented by $ \mathbb{M} = \{1,...,m,...,M\} $. 
According to the time allocation model proposed in subsection \ref{21}, we can classify $ M $ mobile devices and $ U $ UAVs as type-1 and type-2 devices as well as type-1 and type-2 UAVs, respectively. 
Both UAVs and mobile devices are equipped with communication, computation, and energy harvesting units, and UAVs have stronger computing capabilities than mobile devices. 
Mobile devices randomly moving in this region can perform local computing or offload tasks using the energy collected from all APs. 
UAVs driven by the laser emitter provide computing services for mobile devices on the ground, each of which is equipped with a single antenna.
Similar to~\cite{ref2}, we assume that mobile devices is half-duplex, which means that energy harvesting and communication cannot be performed simultaneously.

\indent Furthermore, we model locations of UAVs and mobile devices by three-dimensional Euclidean coordinates. 
To facilitate exposition, the time range is divided into a set of time slots with equal duration $T$. 
In time slot $t$, the coordinate of mobile device $m$ is denoted by $ q_m^t=(x_m^t,y_m^t,0 )$, while that of UAV $u$ is denoted by $ q_u^t=(x_u^t,y_u^t, H )$. 
It is assumed that UAV $u$ flies at uniform speed $ \mathbb{V}_u^t={(q_u^t-q_u^{t-1})}/{T} $ and altitude $ H $. 
Similar to~\cite{ref3,2021ning,2021wang}, we assume that the channel between UAV $u$ and mobile device $m$ is line-of-sight, and the channel gain is given by:
\begin{equation}\label{hmut} 
	\!\!h_{mu}^t= {\xi_0}({d_{mu}^t})^{ - 2} = \frac {\xi_0}{{H^{2} \!+\! {{\left \|{ {{ q_u^t} \!-\! { q_{m}^t}} }\right \|}^{2}}}},\tag{2}
\end{equation}
where $\xi_0$ is the channel power gain at reference distance $d_0 = 1$ m, and $\|.\|$ denotes the Euclidean norm. Variable $d_{mu}^t$ is the horizontal plane distance between UAV $u$ and mobile device $m$ in time slot $t$.

\subsection{Time Allocation Model}\label{21}

\begin{figure}[!t]
	\centering
	\includegraphics[width=3.4in]{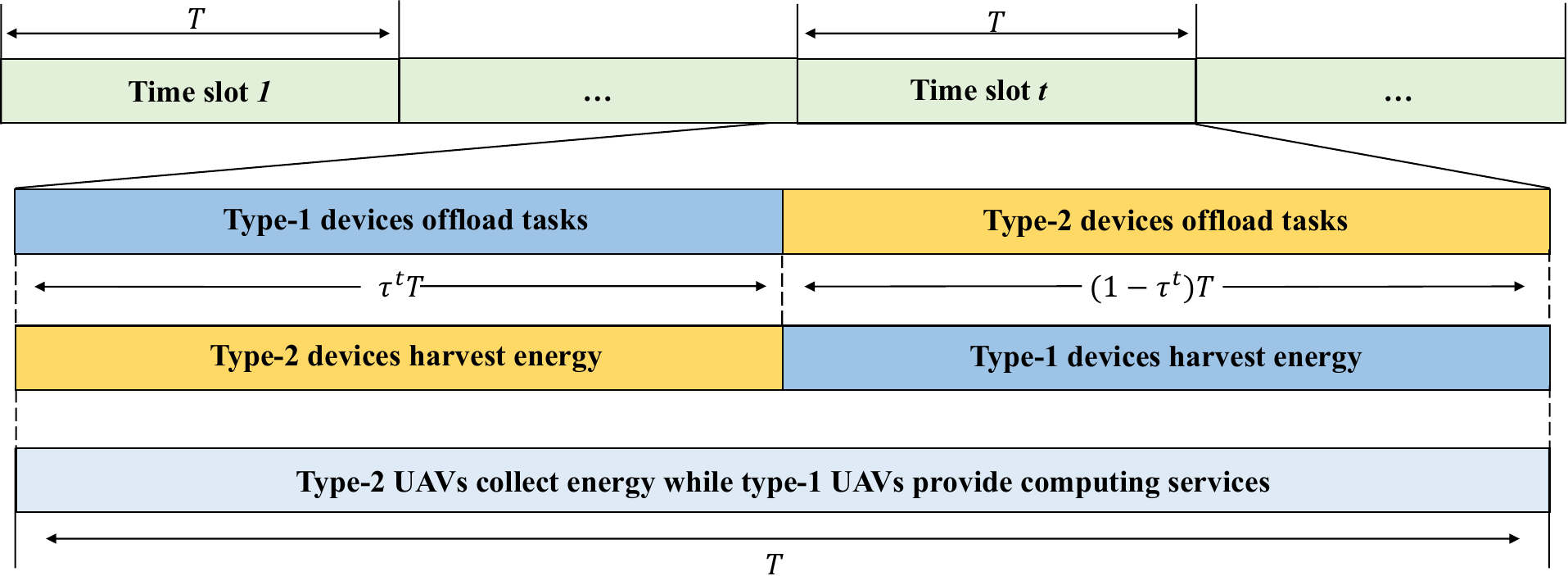}
	\caption{The time allocation model.}
	\label{22}
\end{figure}

\indent Considering half-duplex characteristics of UAVs and mobile devices, we design a time allocation model for the joint scheduling of UAVs and mobile devices. As shown in Fig. \ref{22}, the design principle is as follows:

\indent From the perspective of UAVs, we first divide them into two types, i.e., type-1 and type-2 in each time slot. 
Specifically, type-1 UAVs provide computing services to mobile devices, while type-2 UAVs perform charging operations. 
By optimizing locations of UAVs in each time slot, type-1 and type-2 UAVs can achieve high offloading and charging efficiency, respectively. 
From the perspective of mobile devices, each time slot is divided into two phases, and all devices can be classified into type-1 and type-2 ones. 
Type-1 devices refer to those who perform task offloading operations in the first phase and charging operations in the second phase, while type-2 devices do the opposite.

In time slot $ t $, the scheduling of mobile devices and UAVs is determined by binary variables $ \alpha_m^t $ and $ \beta_u^t $, respectively. 
If the values of $ \alpha_m^t $ and $ \beta_u^t $ are both 1, it means that device $m$ and UAV $u$ are one type-1 mobile device and one type-1 UAV, respectively. Conversely, if both $ \alpha_m^t $ and $ \beta_u^t $ are 0,  device $m$ and UAV $u$ are one type-2 mobile device and one type-2 UAV, respectively. 
The time slot can be divided into two periods by time allocation variable $ \tau^t $, where $ \tau^t \in (0, 1) $. 
In the first period of size $ \tau^t $, type-1 devices offload their computation tasks to type-1 UAVs while type-2 devices collect energy from APs. 
In the second period of size $1 - \tau^t$, type-2 devices offload their tasks to type-1 UAVs while type-1  devices collect energy. 
It is noteworthy that in each time slot, type-1 UAVs always provide computing services to mobile devices, while type-2 UAVs collect energy from the laser transmitter.

Different from existing studies that perform charging or task offloading operations for all devices simultaneously~\cite{ref36}, our time allocation model allows two types of mobile devices to perform different operations in the same period, thus making better utilization of network resources, such as computing resources of type-1 UAVs.
\subsection{Energy Harvesting and Consumption Model}
\subsubsection{Energy Harvesting and Consumption of Mobile Devices}
Since our focus is on joint time scheduling and UAV trajectory design, we adopt the linear energy harvesting model for our system~\cite{ref2}.
Similar to~\cite{ref1}, we view all APs as an integrated virtual transmitter.
Therefore, the harvested energy for device $m$ in time slot $t$ can be expressed by:
\begin{equation}\label{emth}  
	 E_{m}^{t,h}=\varepsilon(1-\alpha_m^t)\widehat{h}_{m}^t\mathcal{P}_0\tau^tT+\varepsilon\alpha_m^t \widehat{h}_{m}^t\mathcal{P}_0(1-\tau^t)T  ,\tag{3}
\end{equation}
where $\varepsilon \in (0,1)$ is the energy conversion efficiency of mobile devices.
Variables $\mathcal{P}_0$ and $ \widehat{h}_{m}^t $ are the transmit power of the virtual transmitter and the downlink channel gain from device $ m $ to the virtual transmitter, respectively.
It should be noted that if device $m$ is type-2, its harvested energy in time slot $t$ is $\varepsilon\widehat{h}_{m}^t\mathcal{P}_0\tau^tT$. 
In case it is of type-1, the collected energy is $\varepsilon \widehat{h}_{m}^t\mathcal{P}_0(1-\tau^t)T$.

We consider partial offloading models, where mobile devices can perform local computation during the entire time slot. 
Specifically, in time slot $t$, the amount of local computation bits of device $m$ can be denoted by  $L_{m}^{t,l}= {f_mT}/{C_1}$.
Symbol $C_1$ represents the number of CPU cycles required to compute each data bit, and variable $f_m$ is the CPU frequency of device $m$. 
In addition, similar to~\cite{ref1,ref35,ref36}, energy consumption of device $ m $ for local computation is described by the form of the cube of its CPU frequency: $E_{m}^{t,l}= k_mf_m^3T$. 
Symbol $k_m$ is a coefficient related to the computation efficiency of device $m$.

Similar to~\cite{ref1,9453824}, the communication between mobile devices and UAVs is based on orthogonal frequency division multiple access technology. 
The amount of offloading data bits of device $m$ to UAV $u$ in time slot $t$ can be represented as:
\begin{align*}\label{lmot}  
	L_{m}^{t,o}&=(\alpha_m^t\tau^t+(1-\alpha_m^t)(1-\tau^t))\mathbb{B}{log}_2\left(1+\frac{P_mh_{mu}^t}{\delta^2}\right)T\\&=
	(1-\tau^t-\alpha_m^t+2\alpha_m^t\tau^t)\mathbb{B}{log}_2\left(1+\frac{P_mh_{mu}^t}{\delta^2}\right)T,\tag{4}
\end{align*}
where variables $\mathbb{B}$ and $\delta^2$ represent the channel bandwidth and the noise power, respectively. Variable $P_m$ denote the transmission power of device $m$. In addition, the amount of its consumed energy for data transmission of device $ m $ in time slot $t$ can be computed by:
\begin{align*}\label{emto}  
	E_{m}^{t,o}&=(\alpha_m^t\tau^t+(1-\alpha_m^t)(1-\tau^t)){TP}_m
	\\&= (1-\tau^t-\alpha_m^t+2\alpha_m^t\tau^t)TP_m.\tag{5}
\end{align*}

We select the UAV with the best channel condition as the target UAV of device $m$. 
The amount of computation bits of device $m$ in time slot $t$ can be represented as $L_m^t=L_{m}^{t,l}+ L_{m}^{t,o}$. Similarly, the amount of its consumed  energy is $E_m^t=E_{m}^{t,l}+E_{m}^{t,o}$.

\subsubsection{Energy Harvesting and Consumption of UAVs}
Similar to~\cite{ref6} and~\cite{ref37}, taking into account the high energy consumption of flying UAVs, we use the laser charging to improve the charging efficiency of UAVs.
Based on the linear energy harvesting model, the harvested energy of UAV $u$ in time slot $t$ can be represented as:
\begin{equation}\label{euth}
	\mathbb{E}_{u}^{t,h}=\epsilon(1-\beta_u^t) g_{\mathrm {ua}}^{t}\mathcal{P}_{1}T, \tag{6}
\end{equation}
where $\epsilon\in (0,1)$ is the energy conversion efficiency for UAVs, and $\mathcal{P}_{1}$ is the transmission power of the laser. 
The laser charging channel from the laser transmitter to UAV $u$ can be calculated by expression $g_{ua}^t=({G\vartheta e^{-\phi d_{ua}^t})/{{(F+\nu d_{ua}^t)}^2}}$, where variables $G$ and $\phi$ represent the area of the telescope (or collector) of the laser receiver and the attenuation coefficient of the channel medium, respectively. 
Additionally, variable $\vartheta$ denotes the optical efficiency of the combined transmission receiver, and variable $d_{ua}^t$ represents the distance between UAV $u$ and the laser transmitter. 
Variable $F$ denotes the size of the initial laser beam, and variable $\nu$ denotes the angle diffraction~\cite{ref6}.

\indent In each time slot, the energy consumption of one type-1 UAV includes both propulsion and computation energy consumption, while that of one type-2 UAV only includes propulsion energy consumption. 
The computation energy consumption of type-1 UAV $u$ in time slot $t$ can be calculated by $\mathbb{E}_{u}^{t,c}= k_u{f_u}^3T$, where variable $k_u$ is the efficiency coefficient of UAV $u$ for processing each data bit, and variable $f_u$ is the CPU frequency of UAV $u$. 
By considering fixed-wing UAVs~\cite{ref6}, the flight energy consumed by UAV $u$ in time slot $t$ is:
\begin{equation}\label{eutf} 
	\mathbb{E}{_{u}^{t,f}}=T \left({\zeta _{1}\| \mathbb{V}_u^t\|^{3}+\frac {\zeta _{2}}{\| \mathbb{V}_u^t\|}}\right), \tag{7}
\end{equation}
where variables $\zeta_{1}$ and $\zeta_{2}$ are fixed parameters related to the specifications of UAVs~\cite{ref6}.
Finally, the energy consumption of UAV $ u $ in time slot $ t $ can be expressed by:
\begin{align*}\label{eut} 
	\begin{split}
		\mathbb{E}_u^t= &\mathbb{E}_{u}^{t,f}+\beta_u^t\mathbb{E}_{u}^{t,c}
		=T \left({\zeta _{1}\| \mathbb{V}_u^t\|^{3}+\frac {\zeta _{2}}{\| \mathbb{V}_u^t\|}} + {\beta_u^tk}_u{f_u}^3\right).
	\end{split} \tag{8}
\end{align*}

\indent As a result, we can define system computation efficiency $ \eta_{CE}^t $ by the ratio of the total amount of computation bits to that of energy consumed by both mobile devices and UAVs in time slot $ t $, i.e.,
\begin{align*}\label{cet}  
	&\eta_{CE}^t=\cfrac{\sum _{m =1}^{M} L_m^t} {\sum _{m =1}^{M} E_m^t + \sum _{u =1}^{U} \mathbb{E}_u^t}.
	&\tag{9}
\end{align*}

\subsection{Problem Formulation}
\indent In time slot $t$, mobile devices and UAVs can be classified into different types based on variables $ \alpha_m^t $ and $ \beta_u^t $, as described in subsection III-B, and their energy harvesting and consumption models are defined in subsection III-C. 
Considering time allocation variable $ \tau^ t $, device scheduling variable $ \alpha_m^t $, UAV scheduling variable $ \beta_u^t $ and UAV location $ q_u^t $, we formulate the computation efficiency maximization problem as follows:

\indent \textit{Objective}:

\indent Our objective is to maximize the average computation efficiency over all time slots, i.e.,
\begin{align*}\label{p1} 
	&\mbox{P1}:\quad \underset { {\tau^ t}, {\alpha_m^t}, {\beta_u^t}, {q_u^t}}{ \max} \lim \limits _{t\to \infty }\frac{1}{t}\sum _{j =1}^{t}\eta_{CE}^j, \tag{10}\\
	&\mbox{s.t.}\quad \ \ {(\alpha}_m^tP_m+k_mf_m^3)\tau^tT\le E_{m}^{t-1,r},m\in \mathbb{M},t\in\{1,2,3....\}, \tag{10a}\\
&\quad \quad \ \ \  E_m^t \le E_{m}^{t-1,r}+E_{m}^{t,h}, m\in \mathbb{M},t\in\{1,2,3....\},\tag{10b}\\
& \quad \quad \ \ \ \mathbb{E}_u^t\le \mathbb{E}_{u}^{t-1,r}+\mathbb{E}_{u}^{t,h}, u\in \mathbb{U},t\in\{1,2,3....\}. \tag{10c}
\end{align*}

\indent \textit{Input}:

\indent UAV set $\mathbb{U} $, mobile device set $ \mathbb{M}$, CPU frequencies $ {\{f_m}\}_{m\in \mathbb{M}} $ and $ {\{f_u}\}_{u\in \mathbb{U}} $,  channel bandwidth $ \mathbb{B} $, noise power $ \delta^2 $,  transmission power $ \mathcal{P}_{1} $,  energy conversion efficiencies $\varepsilon $ and $ \epsilon $, computation efficiency coefficients $ {\{k_m}\}_{m\in \mathbb{M}} $ and $  {\{k_u}\}_{u\in \mathbb{U}} $, device transmission power $ {\{P_m}\}_{m\in \mathbb{M}} $, device residual battery energy $ {\{{E}_{m}^{t-1,r}\}}_{m\in \mathbb{M}} $, UAV residual battery energy $  {\{\mathbb{E}_{u}^{t-1,r}\}}_{u\in \mathbb{U}}$.

\indent \textit{Output}:

\indent 1) Time allocation variable $ \tau^ t\in(0,1)$, $ t\in\{1,2,3....\} $;

\indent 2) Scheduling variable $ \alpha_m^t \in\{0,1\} $ for device $ m$  and $ \beta_u^t\in\{0,1\} $ for UAV $ u$ $, m \in \mathbb{M} $,  $ u \in \mathbb{U} $ and $ t\in\{1,2,3....\} $;

\indent 3) The trajectory of UAVs, i.e., $ q_u^t $ for UAV $ u$,  $ u \in \mathbb{U} $ and $ t\in\{1,2,3....\} $. 

\indent \textit{Constraints}:
 
\indent During the first $\tau^t$ period of time slot $t$, the energy consumed by type-1 device $m$ should be less than its remaining energy in time slot $ t-1 $, as shown in equation (\ref{p1}a).
In time slot $t$, mobile devices must satisfy inequality constraint (\ref{p1}b), where the energy consumed by device $m$ should not exceed the sum of its remaining energy in time slot $t-1$ and the energy harvested in time slot $t$.
Similarly, the energy consumed by UAVs should also be less than the sum of its remaining energy in time slot $t-1$ and the energy harvested in time slot $t$, as shown in inequality constraint (\ref{p1}c).

\indent Based on constraints (\ref{p1}a)-(\ref{p1}c), it can be observed that device scheduling variable $\alpha_m^t$ influences the choice of time allocation variable $\tau^t$, while UAV scheduling variable $\beta_u^t$ and UAV location $q_u^t$ are strongly coupled. 
Therefore, we can derive the following theorem.

\textbf{Theorem 1:} Problem P1 is a mixed integer nonconvex fractional problem that is NP-hard.

Please refer to Appendix A for the proof of
Theorem 1.

\section{Algorithm Design}\label{4}
In this section, to address Problem P1 defined in subsection III.D, we first decompose P1 into two subproblems based on the coupling relationship among variables.
Then, we propose a two-stage alternating optimization algorithm based on multi-task DRL, which can effectively solve the above two subproblems. 
Specifically, in order to support the running of human-centric metaverse applications at mobile devices, we first optimize device scheduling variable $ \alpha_m^t $ and time allocation variable $ \tau^t $ according to the requirements of mobile devices in subsection IV.B, and then optimize UAV scheduling variable $ \beta_u^t $ and location $q_u^t$ based on multi-task DRL in subsection IV.C.

\subsection{Problem Decomposition}
The problem we have formulated as a mixed nonlinear long-term optimization problem belongs to the class of NP-hard problems and involves numerous optimization variables coupled among time slots.
The traditional heuristic algorithm is not suitable for solving Problem P1 due to its properties such as high complexity and inability to adapt to the dynamic network environment.
Moreover, we observe that from the perspective of UAVs, Problem P1 can be decomposed into a series of interrelated subtasks. 
Therefore, we utilize multi-task DRL to solve these subtasks concurrently.

However, it is challenging to solve Problem P1 directly by multi-task DRL approach because of the large number of complex decision variables involved in it.
On the one hand, in our considered scenario of WPMEC assisted by UAVs, device scheduling variable $ \alpha_m^t $ is a binary decision for each device. 
Modeling device scheduling variable $ \alpha_m^t $ as actions, with the presence of merely 20 mobile devices, the dimensionality of the candidate action space will be $ 2^{20} $.
Such a large action space makes it difficult to achieve convergence in algorithm training.
Therefore, device scheduling variable $ \alpha_m^t $ cannot be directly solved by multi-task DRL.
On the other hand, since Problem P1 has four optimization variables, it is difficult to solve them simultaneously by a multi-task DRL algorithm without sacrificing complexity.

To solve the above challenges, we decompose Problem P1 into two subproblems.
Subproblem 1 is to optimize device scheduling and WPT time allocation in time slot $ t $, which can be expressed by:
\begin{align*}\label{p2} 
	&\mbox{P2}:\quad \underset { {\tau^ t}, {\alpha_m^t}}{ \max}\quad   \eta_{CE}^t, \tag{11} \\ &\mbox{s.t.}\quad \ \ \text{inequations}\; (\ref{p1}a)\; \text{and} \; (\ref{p1}b).
\end{align*}

\indent Subproblem 2 is the charging scheduling and trajectory design for UAVs with the optimization objective of maximizing the average computation efficiency, which can be expressed by:
\begin{align*}\label{p3} 
	&\mbox{P3}:\quad \underset {{\beta_u^t}, {q_u^t}}{ \max} \lim \limits _{t\to \infty }\frac{1}{t}\sum _{j =1}^{t}\eta_{CE}^j, \tag{12}\\
	&\mbox{s.t.}\quad\ \ \text{inequation}\; \text{(\ref{p1}c)}.
\end{align*}

We design an efficient heuristic algorithm to solve subproblem 1 in subsection IV. B and design a novel multi-task DRL algorithm to solve subproblem 2 in subsection IV. C. 
It is worth noting that subproblem 2 requires the solution of subproblem 1 as input, and the solution of subproblem 2 also affects the solution of subproblem 1. 
Therefore, the long-term optimization of computation efficiency in Problem P1 can be achieved by alternative optimization between subproblem 1 and subproblem 2.

\subsection{Device Scheduling and Time Allocation}
The focus of Problem P2 is to maximize the computation efficiency by scheduling charging and computation offloading for mobile devices. 
Due to the coupling between device scheduling variable $\alpha_m^t$ and time allocation variable $\tau^t$, Problem P2 is a mixed-integer nonlinear programming problem that is difficult to solve. 
Similar to \cite{ref7} and \cite{ref8}, we consider a device scheduling strategy based on the remaining energy of mobile devices, where devices with high remaining energy above a threshold are classified as type-1 ones, and those with relatively low remaining energy below the threshold are classified as type-2 ones. 
The details are as follows:
\begin{align*}\label{theta} 
	\alpha_m^{t*}=
	\begin{cases}
		1,\quad &{E}_{m}^{t-1,r}>\Theta_{s};\\
		0,\quad &\mbox{else},
	\end{cases}\tag{13}
\end{align*}
where $ \Theta_{s} $ is a threshold and $ {E}_{m}^{t-1,r} $ is the residual energy of device $ m $ in time slot $ t-1 $.
We set threshold $ \Theta_{s} $ based on whether the remaining energy of device $ m $ is sufficient for critical operations (e.g., local computation and task offloading).
Based on constraint (\ref{p1}a), it can be observed that during the first $ \tau^ t $ time period of time slot $ t $, the consumed energy of type-1 device $ m $ should be less than its remaining energy in time slot $ t-1 $.
In addition, if device $ m $ does not conduct WPT charging operation in time slot $ t $, rather it spends the whole time of the time slot for offloading tasks as well as local computation, it can be derived that its maximum energy consumption in time slot $ t $ is $ (k_mf_m^3+P_m)T $.
Therefore, setting threshold $ \Theta_{s} $ to the maximum energy consumption of type-1 device $ m $, denoted as $ \Theta_{s}=(k_mf_m^3+P_m)T$, can make constraint (\ref{p1}a) hold under any value of time allocation variable $ \tau^ t $.

In addition, based on optimal device scheduling variable $ \alpha_m^{t*} $, the number of computation bits, energy consumption and harvesting of device $ m $ can be updated to $ {L_m^t}'$, $ {E_m^t}' $ and $ {E_{m}^{t,h}}' $ by:
\begin{align*}\label{lmt} 
	&{L_m^t}'= \frac{f_mT}{C_1}+
	(1-\tau^t-\alpha_m^{t*} 
	\\
	&\quad\quad +2\alpha_m^{t*}\tau^t)\mathbb{B}{log}_2\left(1+\frac{P_mh_{mu}^t}{\delta^2}\right)T, \tag{14}\\
	&{E_m^t}'= k_mf_m^3T+(1-\tau^t-\alpha_m^{t*}+2\alpha_m^{t*}\tau^t)TP_m, \tag{15}\\
	&{E_{m}^{t,h}}'=\varepsilon(1-\alpha_m^{t*})\widehat{h}_{m}^t\mathcal{P}_0\tau^tT+\varepsilon\alpha_m^{t*} \widehat{h}_{m}^t\mathcal{P}_0(1-\tau^t)T.  \tag{16}
\end{align*}

\indent Furthermore, Problem P2 can be transformed into Problem P2$ ' $ with optimal device scheduling variable $ \alpha_m^{t*} $ for time allocation, which can be expressed by:
\begin{align*}\label{tau*} 
	&\mbox{P2}':\quad \underset { {\tau^ t}}{ \max}\quad \cfrac{\sum _{m =1}^{M} {L_m^t}'} {\sum _{m =1}^{M} {E_m^t}' + \sum _{u =1}^{U} \mathbb{E}_u^t}, \tag{17}\\
	&\mbox{s.t.}\quad \ \ (\alpha_m^{t*} P_m+k_mf_m^3)\tau^tT\le E_{m}^{t-1,r},m\in \mathbb{M}, \tag{17a}\\
		&\quad\quad\quad {E_m^t}' \le E_{m}^{t-1,r}+{E_{m}^{t,h}}',m\in \mathbb{M}.\tag{17b}
\end{align*}
It is a linear fractional problem related to time allocation variable $ \tau^t $.
Particularly, we can obtain explicit results of Problem P2$ ' $, which is described in Theorem 2.

\indent \textbf{Theorem 2:} The optimal value of time allocation variable $ \tau^t $ can be obtained by:
\begin{align*}\label{taut} 
	\tau^{t\ast}=
	\begin{cases}
		\underset { {m}}{ \min}\{\frac{E_{m}^{t-1,r}}{T{(\alpha}_m^{t\ast} P_m+k_mf_m^3)}\},\quad &AD-BC>0;\\
		\underset { {m}}{ \max}\{ \frac{k_mf_m^3+\left(1-\alpha_m^{t\ast}\right)P_m}{\left(1-2\alpha_m^{t\ast}\right)\left(P_m+\varepsilon \widehat{h}_{m}^t\mathcal{P}_0\right)} \\+\frac{-{E_{m}^{t-1,r}{-\varepsilon T\alpha}_m^{t\ast} \widehat{h}_{m}^t\mathcal{P}_0}}{T\left(1-2\alpha_m^{t\ast}\right)\left(P_m+\varepsilon \widehat{h}_{m}^t\mathcal{P}_0\right)} \},\quad &AD-BC<0,
	\end{cases}\tag{18}
\end{align*}
where $ A=\sum_{m=1}^{M}{(2\alpha_m^{t\ast}-1)}\mathbb{B}{log}_2\left(1+{P_mh_{mu}^t}/{\delta^2}\right) $, $ B = \sum_{m=1}^{M}\left({f_m}/{C_1}+(1-\alpha_m^{t\ast})\mathbb{B}{log}_2(1+{P_mh_{mu}^t}/{\delta^2})\right) $, $ C=\sum_{m=1}^{M}{P_m(2\alpha_m^{t\ast}-1)} $ and $ D = \sum_{m=1}^{M}({k}_mf_m^3+P_m(1-\alpha_m^{t\ast}))+\sum_{u=1}^{U}\Big( \beta_u^tk_u{f_u}^3+ \zeta _{1}\| \mathbb{V}_u^t\|^{3}+ {\zeta _{2}}/{\| \mathbb{V}_u^t\|}\Big) $.

Please refer to Appendix B for the proof of Theorem 2.

In addition, we can obtain the relationship among variables $A$, $B$, $C$ and $D$ defined in Theorem 2, which is helpful for us to derive the final results of Problem P2$ ' $ and the corollary is as follows:

\indent \textbf{Corollary 1:} The polarity of $ AD-BC $ is determined by variable $ A $, where $ A=\sum_{m=1}^{M}{(2\alpha_m^{t\ast}-1)}\mathbb{B}{log}_2(1+{P_mh_{mu}^t}/{\delta^2}) $, $ B = \sum_{m=1}^{M}\left({f_m}/{C_1}+(1-\alpha_m^{t\ast})\mathbb{B}{log}_2(1+{P_mh_{mu}^t}/{\delta^2})\right) $, $ C=\sum_{m=1}^{M}{P_m(2\alpha_m^{t\ast}-1)} $ and $ D = \sum_{m=1}^{M}({k}_mf_m^3+P_m(1-\alpha_m^{t\ast}))+\sum_{u=1}^{U}\Big( \beta_u^tk_u{f_u}^3+ \zeta _{1}\| \mathbb{V}_u^t\|^{3}+ {\zeta _{2}}/{\| \mathbb{V}_u^t\|}\Big) $.

Please refer to Appendix C for the proof of Corollary 1.

Based on Corollary 1, we can further analyze and obtain Theorem 3 as follows:

\indent \textbf{Theorem 3:} In time slot $ t $, if the sum of channel capacity of all type-1 devices is greater than that of all type-2 devices, we choose the larger value as the optimal solution for $ \tau^t  $.
Conversely, we choose the smaller value as the optimal solution. 

Please refer to Appendix D for the proof of Theorem 3.

Based on Theorem 2 and Theorem 3,  the optimal solution of Problem P2$ ' $ can be obtained.

\subsection{UAV Scheduling and Trajectory Design}
Since $ \beta_u^t $ is a binary variable and coupled with $ q_u^t $, Problem P3 is a mixed integer nonlinear programming problem, and hard to be solved 
by traditional optimization algorithms or DRL methods for the following reasons:
First, traditional heuristic algorithms cannot detect and adapt to the dynamics of the network environment, resulting in poor performance;
Second, given the large action space of UAVs,  solving Problem P3 by traditional DRL such as policy-based gradients may face slow convergence and large time complexity.

Therefore, we divide the charging scheduling and trajectory design task for all UAVs into $ U $ subtasks.
Specifically, subtask $ u $ aims to complete the charging scheduling and trajectory design for UAV $ u $, where $ u\in \mathbb{U} $.
This significantly reduces the dimensionality of the action space, since only candidate actions for one UAV need to be considered for each subtask.
However, the interactions among the $ U $ subtasks cause them to be not able to develop separate strategies. 
Therefore, we design a multi-task DRL-based scheduling algorithm that can learn the relationships among different subtasks and process them simultaneously to maximize the computation efficiency of the system.
First, we transfer Problem P3 into a Markov Decision Process (MDP), which is presented as follows: 

\begin{figure*}[!t]
	\centering
	\includegraphics[width=6.2in]{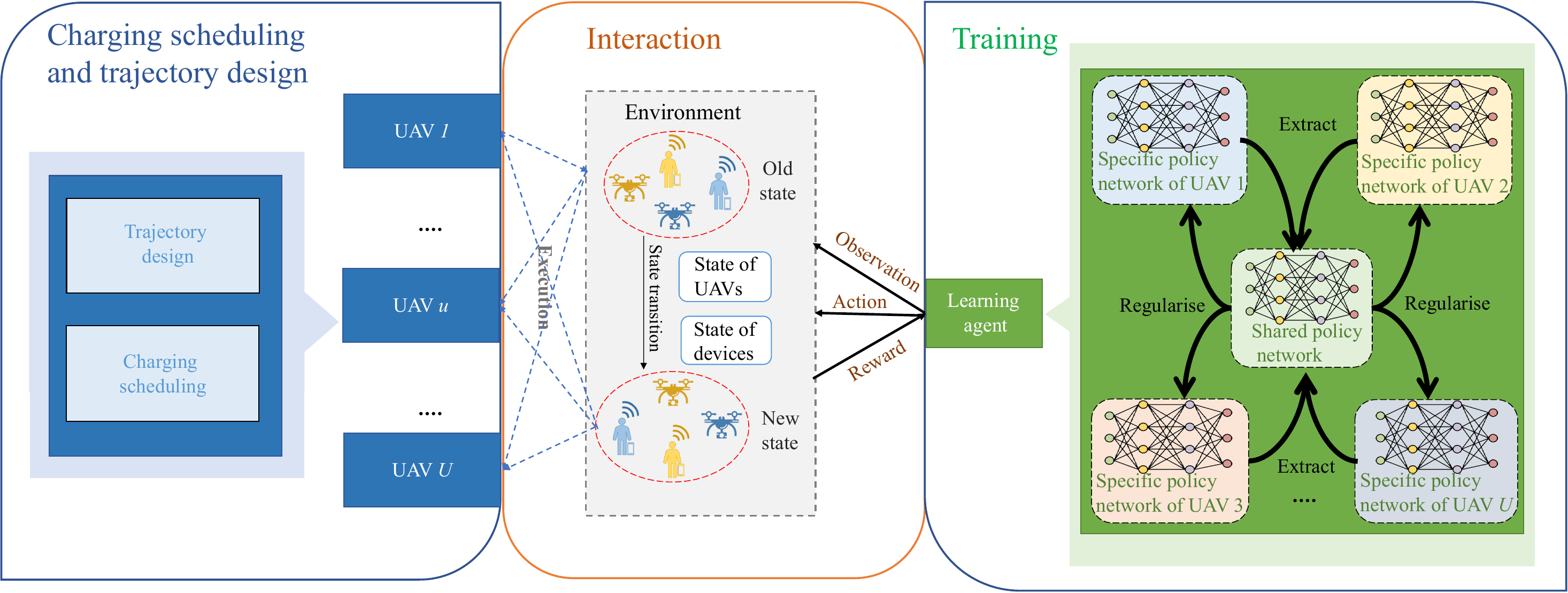}
	\caption{The structure of the designed learning algorithm.}
	\label{33}
\end{figure*}

\subsubsection{MDP Problem Formulation}
\indent The observable Markov game can be formulated by tuple $ (S_u, \mathbb{A}
_u,\mathbb{P}_u, R_u,\gamma) $, where $ u\in \mathbb{U} $.
The learning agent interacts with a dynamic environment that is characterized by action space $ \mathcal{A} \triangleq \{ \mathbb{A}_u, u \in \mathbb{U}\}  $ and state space $ \mathcal{S} \triangleq \{ S_u, u \in \mathbb{U}\}  $, where $ \mathbb{A}_u $ and $ S_u $ are the action space and state space of UAV $ u $, respectively.
The transfer from the current state of UAVs to its next state is based on probability $ \mathcal{P} \triangleq \{ \mathbb{P}_u, u \in \mathbb{U}\}  $.
In addition, symbol $ \gamma \in (0,1) $ refers to the discount coefficient.
We define $ \pi_u $ as the task-specific stochastic policy for charging scheduling and trajectory design of UAV $ u $. 
Considering that charging scheduling and trajectory design among UAVs are interactive, we define $ \pi_0 $ as a shared policy to learn knowledge among multiple UAVs.
As shown in Fig. \ref{33}, during the training process, the action of each UAV is output by a separate policy network, and the shared policy network can extract information from all the policy networks to assist them in formulating a rational policy. 
Thus, each UAV knows not only its private information when selecting an action according to policy $ \pi_u $, but also the additional information about other UAVs (e.g., their coordinates and scheduling variables).
In the following, we describe different elements of the formulated  Markov game.

\indent \textit{States:}
State space $ S_u $ of each UAV composes state space $ \mathcal{S} \triangleq \{ S_u, u \in \mathbb{U}\}  $.
It is observed that $ S_u=\{s_u^t=(q_u^{t-1},\mathbb{E}_{u}^{t-1,r},\{q_m^t,E_{m}^{t-1,r}\}_{m\in\mathbb{M}})\}$.
Set $ \{q_m^t\}_{m\in\mathbb{M}}$ includes coordinates of all devices in time slot $ t $, and $ q_u^{t-1} $ is the coordinate of UAV $ u $ in time slot $ t-1 $. 
Set $ \{E_{m}^{t-1,r}\}_{m\in \mathbb{M}}$ and symbol $ \mathbb{E}_{u}^{t-1,r} $ are the residual energy of all devices and UAV $ u $ in time slot $ t-1 $, respectively.
In addition, the state of all UAVs in time slot $ t $ is denoted as $ s^t=(s^t_1,..., s^t_u,..., s^t_U) $.

\indent \textit{Actions:}
The action space of UAV $ u $ can be expressed by $ \mathbb{A}_u=\{{a_u^t}= (\beta_u^t,q_u^t) \} $. 
Variables $ \beta_u^t $ and $ q_u^t $ are the scheduling variable and the coordinate of UAV $ u $ in time slot $ t $, respectively.
Then, the joint action of UAVs in time slot $ t $ is represented by $ a^t=(a^t_1,... , a^t_u,... , a^t_U)$. 
After executing action $a_u^t$, UAV $ u $ receives reward $ R_u\left(a_u^t\middle| s^t\right) $ and the private state of UAV $ u $ is updated to the new state.

\indent \textit{State Transition:}
Probability $ \mathbb{P}_u\left(s_u^{t+1}\middle| s_u^t,a_u^t\right) $ denotes the state transition probability for UAV $u$, when state $ s_u^t $ shifts to state $ s_u^{t+1} $ by taking action $ a_u^t $.

\indent \textit{Reward:}
The main goal of this work is to maximize the computation efficiency of the system, i.e., to maximize the system computation bits and minimize the total system energy consumption. 
Therefore, UAV $ u $ can obtain the reward according to the defined computation efficiency as follows:
\begin{align*}\label{rut}  
	R_u\left(a_u^t\middle| s^t\right)= \sum _{j =1}^{t}\eta_{CE}^j.  \tag{19} 
\end{align*}
Furthermore, the reward of the learning agent is the sum of rewards of all UAVs, i.e., $ R\left(a^t\middle| s^t\right)=\sum _{u =1}^{U}R_u\left(a_u^t\middle| s^t\right) $.
\indent \textit{Constraint Relaxation:}
During the execution of the multi-task DRL algorithm, it is difficult to satisfy the energy constraint of UAVs.
Thus, we add penalty term $ l_u^t $ to reward $ R_u\left(a_u^t\middle| s^t\right) $, expressed by:
\begin{align*}\label{lut}  
	l_u^t=
	\begin{cases}
		0,\quad &\mathbb{E}_u^t -{\mathbb{E}}_{u}^{t-1,r}-\mathbb{E}_{u}^{t,h} \le 0,\\
		\lambda_u^t,\quad &\mbox{else},
	\end{cases}\tag{20}
\end{align*}
where symbol $ \lambda_u^t $ is an adjustment parameter for UAV $ u $. 
This penalty is to ensure that the energy constraint of UAV  $ u $ can be satisfied, i.e., the sum of the residual energy of UAV $ u $ in time slot $ t-1 $ and the energy collected in time slot $ t $ is higher than the energy consumed in time slot $ t $.
Thus, when the above constraint is satisfied, $ l_u^t = 0 $; otherwise, $ l_u^t = \lambda_u^t $.
Then, the reward is updated by $ R_u^\prime\left(a_u^t\middle| s^t\right)=R_u\left(a_u^t\middle| s^t\right)-l_u^t $, where $ u\in \mathbb{U}$.

\subsubsection{Optimization Problem Formulation}
It is worth noting that for reinforcement learning, maintaining a trade-off between exploration and exploitation is intuitively important to obtain good training results. 
Therefore, to encourage exploration and avoid premature trapping in local optima, we regulate each task-specific policy by $ \gamma $-discount KL difference~\cite{ref9} for shared policies, i.e.,
\begin{align*}\label{eut}  
	\mathbb{E}_{\pi_u}\left[\sum_{t\geq0}\gamma^t\log{\frac{\pi_u\left(a_u^t\middle| s^t\right)}{\pi_0\left(a_u^t\middle| s^t\right)}}\right], u \in \mathbb{U}.\tag{21}
\end{align*}

We need to find shared policy $ \pi_0 $ and task-specific policy $ \{\pi_u\}_{u\in\mathbb{U}}  $ to achieve the long-term minimization of system energy consumption and the long-term maximization of system computation bits, i.e., to solve Problem P3.
In other words, we take maximizing all UAV expected rewards as the optimization objective. 
To avoid local optima, we also introduce $ \gamma $-discount entropy regularization.
As a result, we formulate Problem P4 as follows:
\begin{align*}\label{p4}
	\mbox{P4}:&\max_{\left(\pi_0,\{\pi_u\}_{u\in\mathbb{U}}\right)}\sum_{u=1}^{U} \mathbb{E}_{\pi_u}\sum_{t\geq0}\Big(\gamma^tR_u'\left(a_u^t\middle| s^t\right)-\\&c_{KL}\gamma^t\log{\pi_0\left(a_u^t\middle| s^t\right)}
	-(c_{Ent}-c_{KL})\gamma^t\log{\pi_u\left(a_u^t\middle| s^t\right)} \Big), \tag{22}
\end{align*}
where variables $ c_{KL}, c_{Ent} \ge 0 $ are scalar factors determining the strengths of the KL and entropy regularization.

Then, we introduce two variables, $ \lambda=c_{KL}/(c_{KL}+c_{Ent}) $ and $ \mu=1/(c_{KL}+c_{Ent}) $, and input them into equation (\ref{p4}) and obtain equation (\ref{p41}), which is expressed by:
\begin{align*}\label{p41}
	\mbox{P4}:&\max_{\left(\pi_0,\{\pi_u\}_{u\in\mathbb{U}}\right)}\sum_{u=1}^{U} \mathbb{E}_{\pi_u}\sum_{t\geq0}\Big(\gamma^tR_u'\left(a_u^t\middle| s^t\right)+\\&\frac{\gamma^t\lambda}{\mu}\log{\pi_0\left(a_u^t\middle| s^t\right)}-\frac{\gamma^t}{\mu}\log{\pi_u\left(a_u^t\middle| s^t\right)} \Big) . \tag{23}
\end{align*}
To overcome the exploration and exploration bottleneck, $ \log{\pi_0\left(a_u^t\middle| s^t\right)} $ in equation (\ref{p41}) can be viewed as a reward-shaping term that encourages the selection of actions with high probability under shared policy $ \pi_0 $, while entropy term $ -\log{\pi_u\left(a_u^t\middle| s^t\right)} $ encourages the agent to select other actions.

With respect to the optimization of the objective function, we utilize an alternating maximization procedure, which optimizes over UAV-specific policy $ \{\pi_u\}_{u\in \mathbb{U}} $ given shared policy $ \pi_0 $ and over shared policy $ \pi_0 $ given UAV-specific policies $ \{\pi_u\}_{u\in \mathbb{U}} $, respectively.
Deep Neural Networks (DNNs) are used to parameterize shared policy $ \pi_0 $ and task-specific policies $ \{\pi_u\}_{u\in \mathbb{U}} $.
First, by fixing shared policy $ \pi_0 $, Problem P4 is decomposed into separate maximization problem P4$' $ for each task, which is shown by:
\begin{align*}\label{p42} 
	\mbox{P4}':&\max_{\{\pi_u\}_{u\in\mathbb{U}}} \sum_{u=1}^{U}\mathbb{E}_{\pi_u}\left[ \sum_{t\geq0}R_u^{\prime\prime}\left(a_u^t\middle| s^t\right)-\frac{\gamma^t}{\mu}\log{\pi_u\left(a_u^t\middle| s^t\right)}\Big) \right],\\& u \in \mathbb{U}, \tag{24}
\end{align*}
where the regularization reward is redefined by:
\begin{align*}\label{rut1} 
	R_u^{\prime\prime}\left(a_u^t\middle| s^t\right)=R_u'\left(a_u^t\middle| s^t\right)+\frac{\lambda}{\mu}\log{\pi_0\left(a_u^t\middle| s^t\right)}, u \in \mathbb{U}. &\tag{25}
\end{align*}

Problem P4$ ' $ can be optimized by soft Q learning, which is based on the following ``softened" Bellman updates for states and actions~\cite{ref11}:
\begin{equation}\label{vut} 
	V_u\left(s^t\right)=\frac{1}{\mu}\log{\sum_{a_u^t}\pi_0}\left(a_u^t\middle| s^t\right)\exp \left[ \mu Q_u\left(a_u^t\middle| s^t\right) \right], \tag{26}
\end{equation}
where Q function $Q_u\left(a_u^t\middle| s^t\right)$ can be computed by:
\begin{align*}\label{qut} 
	Q_u\left(a_u^t\middle| s^t\right)&=\gamma^tR_u'\left(a_u^t\middle| s^t\right)+\frac{\gamma^t\lambda}{\mu}\log{\pi_0\left(a_u^t\middle| s^t\right)}\\&+\gamma\sum_{s_u^{t+1}}{\mathbb{P}_u\left(s_u^{t+1}\middle| s_u^t,a_u^t\right)}V_u\left(s^{t+1}\right). \tag{27}
\end{align*}
Through the derivation of equations (\ref{vut}) and (\ref{qut}), it can be derived that optimal policy $ \pi_u $ is a Boltzmann policy at inverse temperature $ \mu $~\cite{ref9}:
\begin{align*}\label{piut} 
	\pi_u\left(a_u^t\middle| s^t\right)&=\pi_0\left(a_u^t\middle| s^t\right)\exp \left[ \mu Q_u\left(a_u^t\middle| s^t\right)-\mu V_u(s^t) \right]
	\\&= \pi_0\left(a_u^t\middle| s^t\right)\exp \left[ \mu A_u\left(a_u^t\middle| s^t\right) \right], \tag{28}
\end{align*}
where $ A_u\left(a_u^t\middle| s^t\right)=Q_u\left(a_u^t\middle| s^t\right)-V_u(s^t) $ is a softened advantage function.

By observing P4$ ' $, it is clear that only $ -{\gamma^t}\log{\pi_u\left(a_u^t\middle| s^t\right)}/\mu  $ is related to $ \pi_u $.
Therefore, optimizing P4$' $ with a fixed shared policy $ \pi_0 $ is equivalent to optimizing Q function $ Q_u\left(a_u^t\middle| s^t\right) $. 
Then, when task-specific policy $ \{\pi_u\}_{u\in\mathbb{U}} $ is fixed, the term in equation (\ref{p41}) depending on shared policy $ \pi_0 $ is:
\begin{align*}\label{pi000} 
	\frac{\lambda}{\mu}\sum_{u=1}^{U}{\mathbb{E}_{\pi_u}}\sum_{t\geq0}\gamma^t\log{\pi_0\left(a_u^t\middle| s^t\right)}. &\tag{29}
\end{align*}
In order to obtain the solution of this problem, we can use stochastic gradient ascent~\cite{ref9}, which precisely leads to an update for extracting all UAV task-specific policies $ \{\pi_u\}_{u\in\mathbb{U}} $ as shared policy $ \pi_0 $.

\subsubsection{The Whole Algorithm}
We construct $ U+1 $ neural networks corresponding to $ U+1 $ optimization parameters in Problem P4, i.e., the shared policy network with optimization parameter $ \theta_0 $, and specific policy networks with $ \{\theta_u\}_{u \in \mathbb{U}}$. 
To learn charging scheduling and trajectory design policies for UAVs, the current state, the action, the next state, the reward and the number of rounds of each UAV task are recorded in memory.
Then, the learning agent extracts the collected data in batches to train those neural networks.
\begin{algorithm}
	\caption{Pseudo-Code of Training Processes for The Learning Agent}\label{alg:alg1}
	\begin{algorithmic}
		\STATE 
		\STATE {\textbf{Input:}} Batch size $ {B} $,  initial shared policy parameter $ \theta_0 $ and specific policies parameters $ \theta_1 $, ..., $ \theta_u $, ..., $ \theta_U $.
		\STATE {\textbf{Output:}} Learned policy $ \pi_0 $, $ \pi_1 $, ..., $ \pi_u $, ..., $ \pi_U $.
		\STATE\hspace{0.2cm}1:\hspace{0.2cm}Initialize initial state and empty memory.
		\STATE\hspace{0.2cm}2:\hspace{0.2cm}\textbf{for} round $ t = 1,2.. $ \textbf{do}
		\STATE\hspace{0.2cm}3:\hspace{0.4cm}\textbf{for} UAV $ u = 1,2, ..., U $ \textbf{do}
		\STATE\hspace{0.2cm}4:\hspace{0.6cm}Obtain $ Q_u $ and $ \pi_0 $ through UAV-specific policy and\\  \hspace{0.95cm} shared policy networks, respectively.
		\STATE\hspace{0.2cm}5:\hspace{0.6cm}$ V_u $ is calculated according to equation (\ref{vut}).
		\STATE\hspace{0.2cm}6:\hspace{0.6cm}$ \pi_u $ is calculated according to equation (\ref{piut}).
		\STATE\hspace{0.2cm}7:\hspace{0.6cm}Select action $ a_u^t $ based on UAV-specific policy $ \pi_u $\\  \hspace{0.95cm} and state $ s^t=(s^t_1,..., s^t_u,..., s^t_U)  $.
		\STATE\hspace{0.2cm}8:\hspace{0.6cm}Obtain device scheduling variable $ \{\alpha_m^{t*}\}_{m\in\mathbb{M}}$ and\\  \hspace{0.95cm} time allocation variable $\tau^{t*}$ by solving Problem P2\\  \hspace{1.1cm}based on action $ a_u^t $ and state $ s^t  $.
		\STATE\hspace{0.2cm}9:\hspace{0.6cm}Obtain state $ s_u^{t+1}$ and reward $ R_u\left(a_u^{t}\middle| s^t\right) $ by\\  \hspace{0.95cm} $ \{\alpha_m^{t*}\}_{m\in\mathbb{M}}$ and $\tau^{t*}$.
		\STATE\hspace{0cm}10:\hspace{0.6cm}Store experience $ \{s_u^t,a_u^t,s_u^{t+1},R_u\left(a_u^t\middle| s^t\right),t\} $ in the\\  \hspace{0.95cm} replay memory.
		\STATE\hspace{0cm}11:\hspace{0.6cm}Sample a batch of state transition\\  \hspace{0.95cm} $ \{s_u^t,a_u^t,s_u^{t+1},R_u\left(a_u^t\middle| s^t\right),t\} $ with size $ B $.
		\STATE\hspace{0cm}12:\hspace{0.6cm}Derive gradient $ \nabla_{\theta_u} $ for all of the episodes in the\\  \hspace{0.95cm} batch based on equation (\ref{lutt}).
		\STATE\hspace{0cm}13:\hspace{0.6cm}Train UAV-specific policy network $ u $ with gradient\\  \hspace{0.95cm} $ \nabla_{\theta_u} $ by Adam  optimizer.
		\STATE\hspace{0cm}14:\hspace{0.4cm}\textbf{end for}
		\STATE\hspace{0cm}15:\hspace{0.4cm}Sample a batch of state transition\\  \hspace{0.75cm} $ \{s^t,a^t,s^{t+1},R\left(a^t\middle| s^t\right),t\} $ with size $ B $.
		\STATE\hspace{0cm}16:\hspace{0.4cm}Derive gradient $ \nabla_{\theta_0} $ for all of the episodes in the batch\\  \hspace{0.75cm}
		by equation (\ref{l0tt}).
		\STATE\hspace{0cm}17:\hspace{0.4cm}Train the shared policy network with gradient $ \nabla_{\theta_0} $ by\\  \hspace{0.75cm} Adam optimizer.
		\STATE\hspace{0cm}18:\hspace{0.2cm}\textbf{end for}
	\end{algorithmic}
	\label{alg1}
\end{algorithm}

\indent We train the $ U+1$ networks by two steps.
First, we solve Problem P4$' $ with fixed $ \pi_0 $ and obtain learning values $ \{\theta_u\}_{u \in \mathbb{U}}$ by training specific policy networks.
We set the loss function of specific policy network $ u $ by:
\begin{align*}\label{lutt} 
	L(\theta_u)=\mathbb{E}_{\pi_u}\left[\Big(Q_u\left(a_u^t\middle| s^t\right)-\widetilde{Q_u}\left(a_u^t\middle| s^t\right)\Big)^2 \right], u\in \mathbb{U}, &\tag{30}
\end{align*}
where $ \widetilde{Q_u}\left(a_u^t\middle| s^t\right) $ and $ Q_u\left(a_u^t\middle| s^t\right) $ are the real Q value of the output of specific policy network $ u $  and the expected Q value computed by equation (\ref{qut}), respectively. After that, we train shared policy $ \pi_0 $ to learn parameter $ \theta_0 $ with fixed policy $ \pi_u $, with the purpose of solving Problem P4. 
The corresponding loss function is set by:
\begin{align*}\label{l0tt} 
	L(\theta_0)=\sum_{u=1}^{U}{\mathbb{E}_{\pi_u}\sum_{t\geq0}\gamma^t\log{\pi_0\left(a_u^t\middle| s^t\right)}}. &\tag{31}
\end{align*}
Finally, we use the Adma optimizer~\cite{ref9} to train the above neural networks.
The training process for the learning agent can be found in Algorithm 1.

\begin{algorithm}
	\caption{Pseudo-Code of MURAL}\label{alg:alg2}
	\begin{algorithmic}
		\STATE 
		\STATE {\textbf{Input:}}$ {\{f_m}\}_{m\in \mathbb{M}} $, $ {\{f_u}\}_{u\in \mathbb{U}} $, $ \mathbb{B} $, $ N $, $ \delta^2 $, $ \mathcal{P}_{1} $, $\varepsilon   $, $ \epsilon $, $ {\{k_m}\}_{m\in \mathbb{M}} $, $  {\{k_u}\}_{u\in \mathbb{U}} $, $ {\{P_m}\}_{m\in \mathbb{M}} $, $ {\{{E}_{m}^{t-1,r}\}}_{m\in \mathbb{M}} $, $  {\{\mathbb{E}_{u}^{t-1,r}\}}_{u\in \mathbb{U}}$ and tolerance errors $ \psi_1 $, $  \psi_2$ and $  \psi_3$.
		\STATE {\textbf{Output:}} Average computation efficiency $\frac{1}{t}\sum _{j =1}^{t}\eta_{EE}^j $.
		\STATE\hspace{0.2cm}1:\hspace{0.2cm}Initialize UAV scheduling variable $ \{\beta_u^{0*}\}_{u\in\mathbb{U}} $ and UAV\\  \hspace{0.55cm} location $ \{q_u^{0*}\}_{u\in\mathbb{U}} $.
		\STATE\hspace{0.2cm}2:\hspace{0.2cm}\textbf{for} time slot $ t = 1,2... $ \textbf{do}
		\STATE\hspace{0.2cm}3:\hspace{0.4cm}Get device scheduling variables $ \{\alpha_m^{t*}\}_{m\in\mathbb{M}}$ based on\\  \hspace{0.75cm} the residual energy of devices.
		\STATE\hspace{0.2cm}4:\hspace{0.4cm}\textbf{for} round $ i= 1,2...$ \textbf{do}
		\STATE\hspace{0.2cm}5:\hspace{0.6cm}$ \beta_u^{t,0}=\beta_u^{(t-1)*} $, $ q_u^{t,0}=q_u^{(t-1)*} $.
		\STATE\hspace{0.2cm}6:\hspace{0.6cm}Obtain time allocation variable $ \tau^{t,i} $ by solving\\  \hspace{0.95cm} Problem P2 based on $ \{\alpha_m^{t*}\}_{m\in\mathbb{M}}$,  $\{\beta_u^{t,(i-1)}\}_{u\in\mathbb{M}}$ and\\ \hspace{1.1cm}$\{q_u^{t,(i-1)}\}_{u\in\mathbb{M}}$.
		\STATE\hspace{0.2cm}7:\hspace{0.6cm}Solve Problem P3 by the learned policy (obtained\\  \hspace{0.95cm} by Algorithm 1) for given $ \{\alpha_m^{t*}\}_{m\in\mathbb{M}}$ and $ \tau^{t,i} $.
		\STATE\hspace{0.2cm}8:\hspace{0.6cm}Update $ i = i+1 $, $ \{\beta_u^{t,i}\}_{u\in\mathbb{U}} $ and $ \{q_u^{t,i}\}_{u\in\mathbb{U}} $.
		\STATE\hspace{0.2cm}9:\hspace{0.6cm}\textbf{If} $ \sum_{i^{'}=1}^{i}\|\beta_u^{t,i^{'}}-\beta_u^{t,(i^{'}-1)} \| \le \psi_1 $, $\sum_{i^{'}=1}^{i}\|q_u^{t,i^{'}}-$\\  \hspace{1.1cm}$  q_u^{t,(i^{'}-1)} \| \le \psi_2$ and $ \sum_{i^{'}=1}^{i}\|\tau^{t,i^{'}}-\tau^{t,(i^{'}-1)} \| \le \psi_3 $
		\STATE\hspace{0.0cm}10:\hspace{1.1cm}$\beta_u^{t*}= \beta_u^{t,i} $,  $q_u^{t*}= q_u^{t,i} $, $\tau^{t*}=\tau^{t,i}$.
		\STATE\hspace{0.0cm}11:\hspace{1.1cm}Break.
		\STATE\hspace{0.0cm}12:\hspace{0.6cm}\textbf{end if}
		\STATE\hspace{0.0cm}13:\hspace{0.4cm}\textbf{end for}
		\STATE\hspace{0.0cm}14:\hspace{0.4cm}Update $ t=t+1 $.
		\STATE\hspace{0.0cm}15:\hspace{0.2cm}\textbf{end for}
	\end{algorithmic}
	\label{alg2}
\end{algorithm}

Overall, the execution process of the designed algorithm, MURAL, is as follows:
First, in time slot $ t $, we obtain optimal device scheduling variable $ \{\alpha_m^{t*}\}_{m\in\mathbb{M}}$ according to the device scheduling policy based on the remaining energy of mobile devices.
Second, we design both heuristic and multi-task DRL algorithms.
By alternating iterations of heuristic algorithm and multi-task DRL algorithm, we obtain optimal time allocation variable $ \tau^{t*}$, optimal UAV scheduling variable $\{\beta_u^{t*}\}_{u\in\mathbb{U}}$ and optimal UAV coordinate $\{q_u^{t*}\}_{u\in\mathbb{U}}$ in time slot $ t $.
In addition, the execution process of MURAL algorithm can be found in Algorithm 2.

\subsubsection{Algorithm Complexity Analysis}
We perform theoretical analysis of the complexity of the designed multi-task DRL-based algorithm (Algorithm 1) and MURAL algorithm (Algorithm 2) as follows. \\
\indent \textbf{Theorem 4:} The complexity of Algorithm 1 is $ \mathcal {O}\Big(\vphantom{\sum \limits _{i=1}^{\Re -1}} L \times E \times B \times \Big( \sum \limits _{u=1}^{U} \sum \limits _{i=1}^{\Re -1} \imath_{u,i}\times \imath _{u,i+1} + \sum \limits _{j=1}^{\Im -1} \imath _{0,i}\times \imath _{0,i+1} \Big) \Big) $. 
Variables $ \imath_{u,i} $ and $ \imath_{0,i} $ represent the number of neurons at layer $ i $ of UAV-specific policy network $ u $ and the shared policy network, respectively.
Variables $ \Re $ and $ \Im $ are the number of fully connected layers of each UAV-specific policy network and the shared policy network. 
Meanwhile, variables $ E $, $ L $ and $ B $ refer to the number of training episodes, the number of time slots per episode and the training batch size, respectively.

Please refer to Appendix E for the proof of Theorem 4.

\indent \textbf{Theorem 5:} The complexity of Algorithm 2 is $ \mathcal{O}\Big(\sum_{t=1}^{N} L_t\Big(\sum \limits _{u=1}^{U} \sum \limits _{i=1}^{\Re -1}  \imath_{u,i}\times \imath _{u,i+1} +M\Big)\Big) $.
Symbol $N$ denotes the number of time slots required for the execution of MURAL algorithm. Meanwhile, $ L_t,\;t\in(0,N) $, denotes the number of iterations required to obtain the optimal scheduling results for both UAVs and mobile devices in time slot $t$. 

Please refer to Appendix F for the proof of Theorem 5.

\section{Performance Evaluation}\label{5}

\subsection{Simulation Setup}
\begin{figure}[!t]
	\centering
	\includegraphics[width=2.8in]{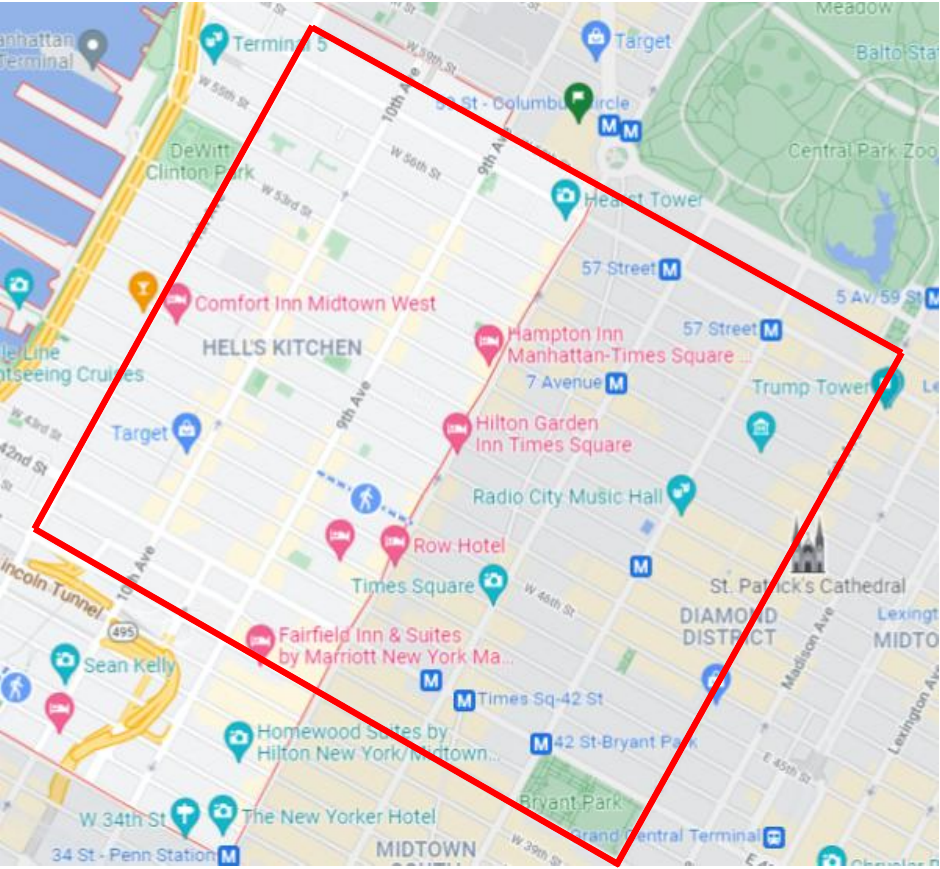}
	\caption{The city map of Manhattan.}
	\label{map}
\end{figure}

In order to evaluate the performance of MURAL, we conducted a set of experiments by Python 3.9 and Pytorch 1.8.1.
As shown in Fig. \ref{map}, we verify the performance utilizing a city map of Manhattan.
We choose an area of 1km $ \times $1km indicated by the red line, where the laser emitter is positioned in the area center, and APs are uniformly distributed in this area.
According to Manhattan mobility model~\cite{ref1}, pedestrians holding mobile devices are treated as mobile devices.
The settings of simulated parameters are based on~\cite{ref1,ref37,ref6}, and can be found in Table \ref{canshu2}.

Because existing studies fail to take into account joint task scheduling and trajectory optimization for multiple UAVs and multiple devices in UAV-assisted WPMEC networks, we evaluate the proposed algorithm, MURAL, versus the following four solutions:
\begin{itemize}{}{}
	\item{TEAM~\cite{ref15}: It is a multi-agent DRL-based scheduling algorithm for multiple UAVs, tending to minimize energy consumption of UAVs by charging scheduling and trajectory design.}
	\item{SCR~\cite{ref16}:  It is a DRL-based resource allocation algorithm to improve system throughput. 
		We apply it in our system. 
		For the time allocation variable, we use a DNN to optimize it. 
		For others, we use a Lagrangian duality method to solve them.}
	\item{No Scheduling of Devices (NSD): UAV charging scheduling and trajectory design are set the same with our algorithm, while without device charging scheduling.}
	\item{Offloading Only (OO): All tasks are offloaded to UAVs for processing without local computation.}
\end{itemize}
\subsection{Simulation Results}

\subsubsection{Learning Performance Evaluation}
\begin{figure}[!t]
	\centering
	\includegraphics[width=3in]{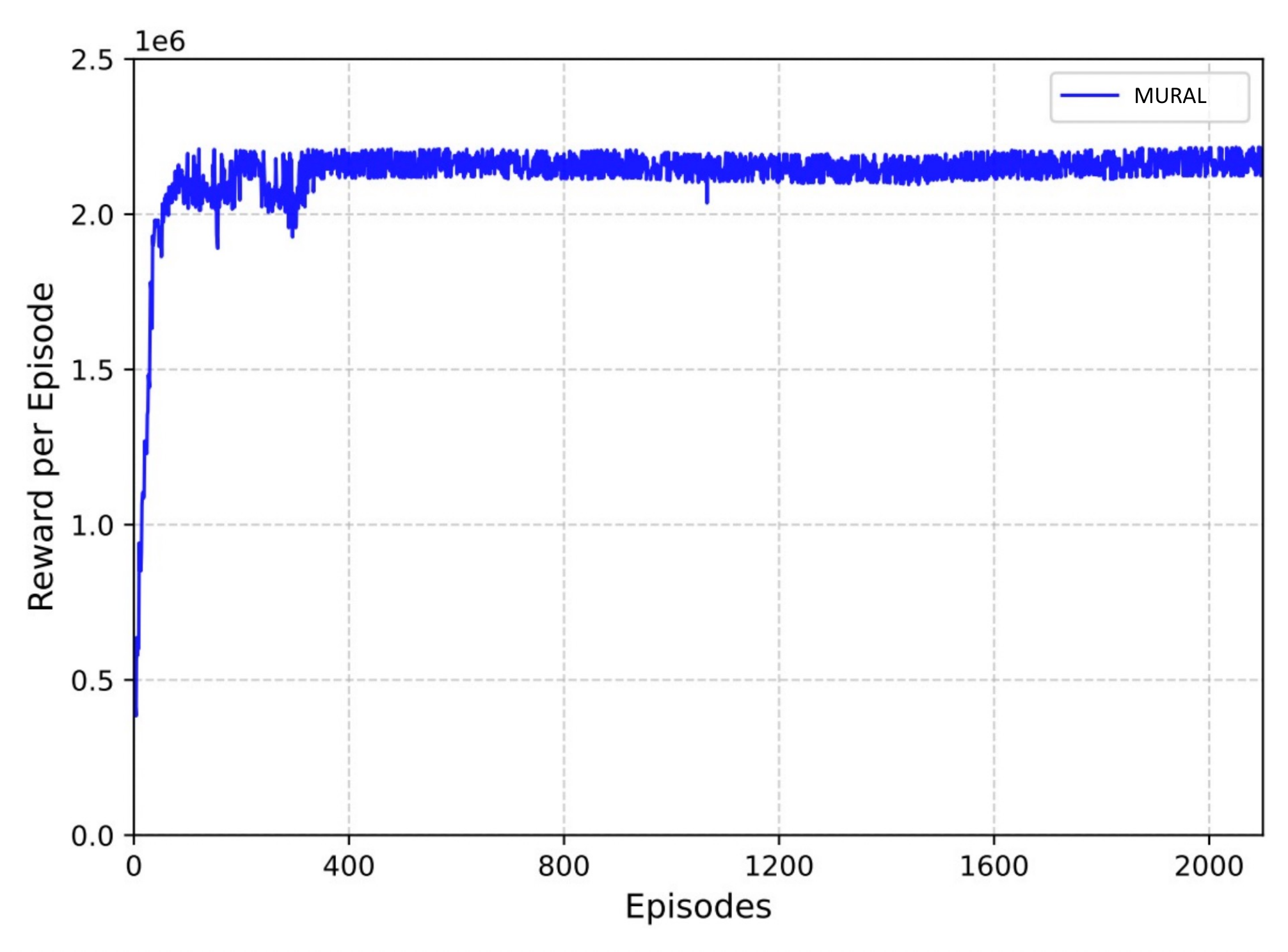}
	\caption{Reward per episode in MURAL (the number of UAVs, APs and mobile devices is 4, 4 and 40, respectively).}\label{train}
\end{figure}

The training curve of MURAL in Fig. \ref{train} is obtained by deploying 4 APs and 4 UAVs to serve 40 devices.
Note that the reward is set according to the objective function defined in equation (\ref{rut}), which includes the total number of computation bits of devices, energy consumption of both devices and UAVs.
It is clearly shown that the obtained reward for each episode remains under 2.5$ \times  {10}^6 $ at the beginning and increases from episode 10, iterating to convergence in episode 380.
First, at the beginning, the learning agent selects random actions to explore the environment and its dynamics.
Second, the shared policy network and UAV-specific policy networks are trained by all the experiences learned from exploration steps to optimally serve mobile devices.
These two steps allow the learning agent to avoid different penalties and optimize placements of UAVs.
This can significantly increase the rewards obtained by UAVs.
Furthermore, it is worth noting that the designed MURAL algorithm has good convergence performance, with rewards gradually achieving convergence when the training episodes reach about 380, where dynamic environmental features and strategy exploration lead to fluctuations in rewards.
\subsubsection{Performance Under Different Numbers of Mobile Devices}

\begin{table}[!t]
	\centering
	\caption{{Simulation Parameters}}
	\label{canshu2}
	\begin{tabular}{lc}
		\toprule[0.8pt]
		{Parameter description }                                                                           & {{Value}}  \\ 
		\toprule[0.8pt]
		The number of UAVs                                                                  & \{4,5,6,7,8\}              \\
		The number of APs                                                                   & \{4,5,6,7,8,9\}            \\
		The number of mobile devices                                                        & \{20,40,60,80,100\}        \\
		The length of each time slot                                                        & 1s                         \\
		The bandwidth                                                                       & 1MHz                       \\
		Transmission power of mobile devices       & 0.1W                       \\
		CPU frequency of each mobile device        & 500MHz                     \\
		CPU frequency of each UAV                                                           & 2.5 GHz                    \\
		Required CPU of tasks                                                                   & 50 cycles/bit              \\
		Transmission power of APs                                                   & 60W                        \\
		Transmission  power of the laser emitter        & 200W                       \\
		Energy harvesting efficiency of mobile\\  devices                                                        & 0.5                        \\
		Energy harvesting efficiency of UAVs                                                  & 0.8                        \\
		The flight altitude of the each UAV                                                      & 10m                        \\
		The maximum moving speed of mobile\\ devices & 20 km/h                    \\
		\begin{tabular}[c]{@{}l@{}}The learning rate for training models\end{tabular}      & 0.001                      \\
		The batch size                                                                      & 128                        \\ 
		\toprule[0.8pt]  
	\end{tabular}
\end{table}

\begin{figure*}[!t]
	\centering
	\includegraphics[width=7in]{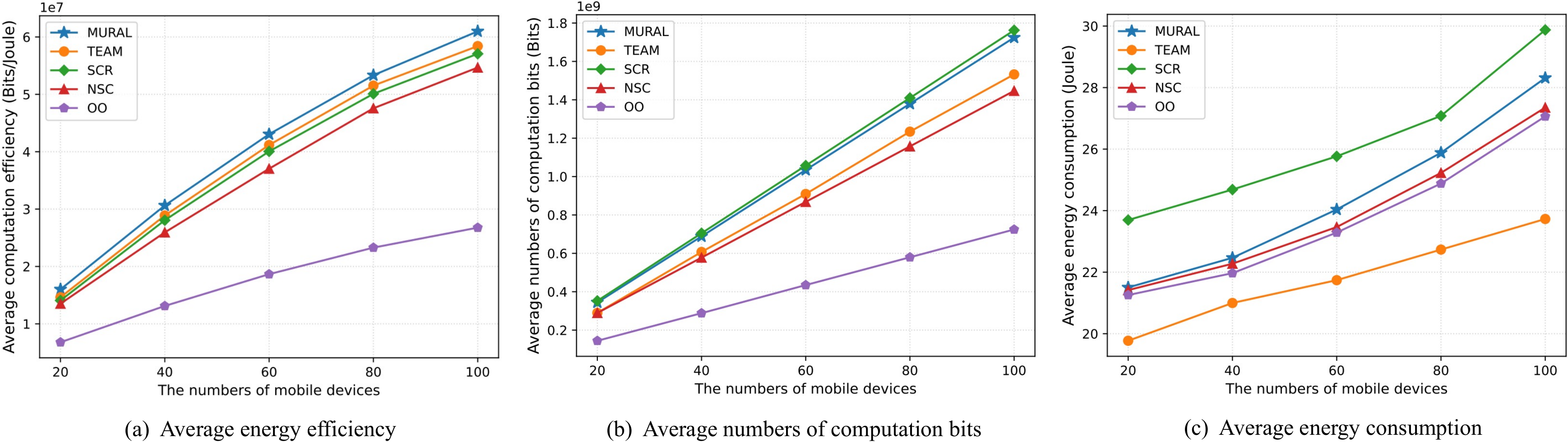}
	\caption{Performance with different numbers of mobile devices (the number of UAVs and APs is 4 and 4, respectively).}
	\label{fang1}
\end{figure*}

Fig. \ref{fang1}a shows the performance of average computation efficiency as defined by equation (\ref{p1}) for TEAM, SCR, NSD, OO and the proposed algorithm, MURAL, under different numbers of mobile devices.
We can discover that the average computation efficiency of MURAL is higher than the other four algorithms.
It is because our algorithm aims to maximize the average computation efficiency by joint resource scheduling and UAV trajectory control.
TEAM algorithm uses multi-agent DRL to determine scheduling variables and trajectories of UAVs, aiming at minimizing the UAV's energy consumption.
Therefore, its performance on average computation efficiency is not as good as that of MURAL algorithm.

SCR is a DRL-based time allocation and resource scheduling algorithm that first fixes the energy collection period and device scheduling, and then solves the charging scheduling and trajectory design problem for UAVs by Lagrangian duality method.
Then, the energy collection period and  device scheduling are derived by an approximation algorithm.
However, only the maximum computation bits per device is considered in the optimization process, without considering the long-term computation efficiency.
Therefore, its performance on average computation efficiency  is close to that of TEAM.

\indent Although NSD shares the same UAV charging scheduling and trajectory design with our algorithm, it uses a fixed energy collection period and does not consider computation scheduling for devices. 
Since it does not fully utilize computational resources in the system, its average computation efficiency is lower than that of MURAL, TEAM and SCR algorithms. 
Finally, the OO algorithm only relies on computation offloading from devices to UAVs without considering the device's local computational capability, which leads to lower performance than the other four algorithms.
When the number of mobile devices increases, the average computation efficiency of all algorithms increases.
The reason for this is that more mobile devices generate more computation  tasks.
Compared to energy consumption of devices and UAVs, the flight energy consumption caused by UAVs to reach the optimal communication location is relatively large and only varies slightly with the increasing number of devices.
Thus, the average computation efficiency of the five algorithms increases with the number of devices.

The performance of average numbers of computation bits under different amounts of mobile devices is shown in Fig. \ref{fang1}b.
Average numbers of computation bits are defined by $\lim \limits _{t\to \infty }\frac{1}{t}\sum _{j =1}^{t}\sum _{m =1}^{M} L_m^j $.
We note that the average numbers of computation bit of MURAL is higher than those of TEAM, NSD and OO algorithms, while lower than that of SCR algorithm.
That's because SCR algorithm solves a deterministic optimization problem by DRL to maximize the computation bits.
Whereas, our work aims to maximize average computation efficiency by a designed multi-task learning algorithm.
In addition, TEAM, which focuses on reducing the energy consumption of UAVs based on multi-agent DRL, and NSD, which does not conduct device charging scheduling, have moderate performance.
Finally, the OO algorithm, which considers only computation offloading, has the worst performance, because it ignores the local computing power of mobile devices.

The performance of the average energy consumption for different amounts of mobile devices is illustrated in Fig. \ref{fang1}c, which is defined by  $\lim \limits _{t\to \infty }\frac{1}{t}\sum _{j =1}^{t}({\sum _{m =1}^{M} E_m^j + \sum _{u =1}^{U} \mathbb{E}_u^j}) $.
We can find that the average energy consumption of TEAM algorithm is the lowest, because it always chooses UAV location that have the minimum energy consumption of the flight.
In order to maximize the computation bit, SCR algorithm usually selects UAV locations with better communication conditions, resulting in higher UAV flight energy consumption compared to the other four algorithms. 
Both NSD and OO algorithms have lower average energy consumption than our proposed algorithm, due to merely considering computation offloading without local processing.
Therefore, the average energy consumption of MURAL is less than that of SCR algorithm and more than that of TEAM, NSD and OO.

\subsubsection{Performance Under Different Numbers of APs}

\begin{figure*}[!t]
	\centering
	\includegraphics[width=7in]{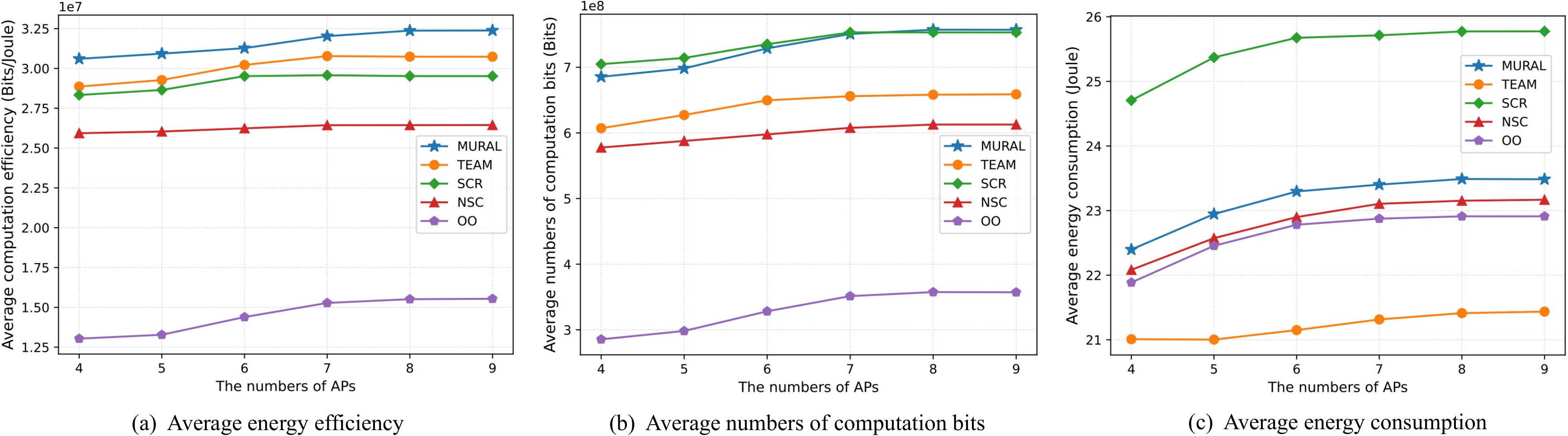}
	\caption{Performance with different numbers of APs (the number of UAVs and mobile devices is 4 and 40, respectively).}
	\label{fang2}
\end{figure*}

Fig. \ref{fang2} shows the impact of different numbers of APs on system performance.
The performance of the average computation efficiency under different numbers of APs is shown in Fig. \ref{fang2}a.
When the number of APs increases, the average computation efficiency increases as well.
It can be found that the average computation efficiency of all algorithms tends to be stable when the number of APs is 7. 
This is because the number of APs mainly affects the energy collection process of devices. 
When the number of APs is 7, most devices can collect enough energy to meet the energy demand in each time slot. 
Therefore, when the maximum CPU frequency and the maximum transmission power of the device are fixed, the number of APs can be chosen by 7 to reach good system performance under current settings.

Fig. \ref{fang2}b and Fig. \ref{fang2}c show the trend of average numbers of computation bits and average energy consumption with different numbers of APs, respectively.
Similarly, it can be found that the average numbers of computation bits and the average energy consumption gradually increase with the increasing number of APs when it is less than 7. 
This is because with the increasing number of APs, the amount of energy collected by devices for local computation and task offloading also increases. 
Thus, the computation bits increase with the number of APs. 

Likewise, with the increase in the collected energy of devices, its computation and transmission energy consumption gradually increases, and UAVs also needs to consume more energy for more offloaded data. 
Therefore, the energy consumption of each algorithm is proportional with the amount of APs.
In addition, the average numbers of computation bits and the average energy consumption tend to be stable when the number of APs is bigger than 7.
This is because when the number of APs is bigger than or equal to 7, the majority of devices can collect enough energy to support their operations.
Thus, when the number of APs changes from 7 to 9, the variation of the average energy consumption and average numbers of computation bits of the system tends to be smooth out.

\subsubsection{Performance Under Different Numbers of UAVs}
\begin{figure*}[!b]
	\centering
	\includegraphics[width=7in]{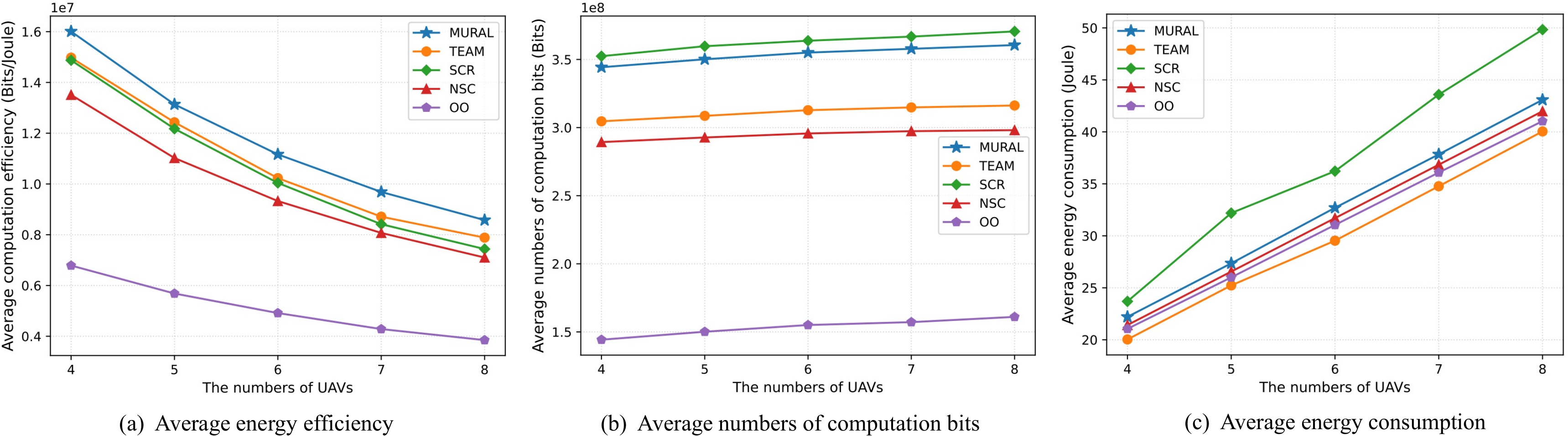}
	\caption{Performance with different numbers of UAVs (The number of APs and mobile devices is 4 and 40, respectively).}
	\label{fang3}
\end{figure*}
The performance of average computation efficiency under different numbers of UAVs is illustrated in Fig. \ref{fang3}a, while Fig. \ref{fang3}b and Fig. \ref{fang3}c show the trend of average numbers of computation bits and average energy consumption under different amount of UAVs, respectively.
By observing Fig. \ref{fang3}a, it can be found that the performance of all algorithms is similar with that shown in Fig. \ref{fang1}a, i.e., MURAL has the lowest average computation efficiency, and TEAM algorithm is the next, while SCR, NSD and OO are less efficient.
By observing Fig. \ref{fang3}b, it can be noticed that the average numbers of computation bits increase gradually when the number of UAVs increases. 
The reason is that, devices have more options to offload tasks with more UAVs.
In addition, we can find that the performance of all algorithms in Fig. \ref{fang3}c is also similar to that of Fig. \ref{fang1}c, i.e., the SCR algorithm has the highest average energy consumption, and MURAL algorithm is the second, and NSD, OO, and TEAM algorithms are the lowest.

Moreover, the flight energy consumption of one UAV is usually higher than computation and transmission energy consumption of devices (e.g., 5 W for one UAV flying in one time slot and 0.0625 W for one device in total).
Therefore, as shown in Fig. \ref{fang3}c, the average energy consumption of each algorithm increases significantly with the increasing number of UAVs.
It can be observed that when the amount of UAVs is increasing, the average computation efficiency is decreasing, as shown in Fig. \ref{fang3}a. 
Compared to the average energy consumption, the average numbers of computation bits fluctuate less with the change in the number of UAVs. 
Thus, it reflects the fact that the average computation efficiency decreases when the amount of UAVs increases.

\subsubsection{The Performance of Execution and Convergence Time}
\begin{figure}[!t]
	\centering
	\includegraphics[width=3.5in]{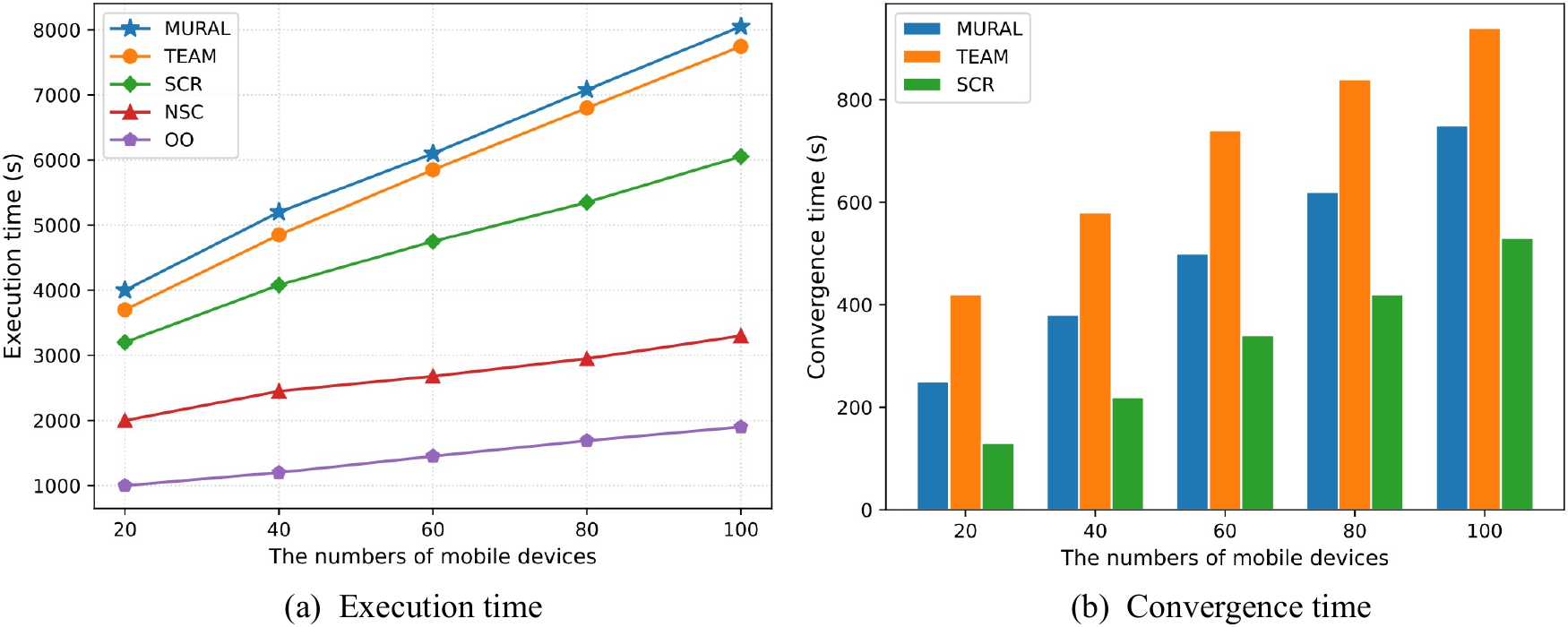}
	\caption{ Performance of execution and convergence time.}
	\label{pic10}
\end{figure}

The execution time of the five algorithms is illustrated in Fig. \ref{pic10}a.
MURAL contains $U+1$ neural networks and trains them based on the collected data batches, so it has the longest execution time.
Next, due to the fact that TEAM also requires training multiple neural networks to predict the trajectories of UAVs, its execution time is only lower than that of our proposed algorithm.
By considering that SCR algorithm only uses 1 DNN, it results in lower execution time than MURAL and TEAM algorithms.
However, NSD and OO algorithms do not have the training process.
The NSD algorithm does not consider computation scheduling of devices, while the OO algorithm only takes into account the offloading of tasks to the UAV and does not concern charging scheduling and time allocation of mobile devices.
The execution time of the five algorithms also becomes long when the number of mobile devices increases.
This is due to the fact that when the number of devices grows, the determination of device scheduling and time allocation variables becomes more complex, and the decision of charging scheduling and location coordinates of UAVs becomes more difficult and thus takes more time.

Fig. \ref{pic10}b shows the convergence time of MURAL, TEAM and SCR algorithms.
We can observe that the convergence time of the three algorithms becomes longer when the number of mobile devices increases.
That's because more devices result in more complex task scheduling, time allocation and UAV trajectory design, with correspondingly longer convergence time.
Furthermore, we can note that the convergence speed of MURAL is between those of TEAM and SCR algorithms.
This is because TEAM algorithm is based on multi-agent DRL, which requires distributed training of multiple agents to optimize trajectories of UAVs. 
As a result, it has the slowest convergence speed. 
In addition, our proposed MURAL algorithm is based on multi-task DRL, which requires only one learning agent to learn charging scheduling and trajectory design of multiple UAVs simultaneously.
Therefore, MURAL algorithm has a faster convergence speed than TEAM algorithm.
Finally, SCR algorithm uses only one DNN to predict time allocation variable, resulting in a faster convergence speed than MURAL and TEAM algorithms.

\section{Conclusion}\label{6}
In this paper, we considered a UAV-assisted WPMEC system that can provide support for human-centric metaverse applications running on mobile devices, and proposed a solution to enable joint charging time allocation, computation task scheduling and trajectory design for both UAVs and mobile devices, with the objective of maximizing the average computation efficiency of the system in a long-term perspective.
First, we formulated the computation efficiency maximization problem by considering energy constraints of mobile devices and UAVs. 
Then, we decomposed it into two subproblems, i.e., optimizing time allocation and charging scheduling of mobile devices, and optimizing charging scheduling and location coordinates of UAVs.
After that, we designed a two-stage alternating optimization algorithm based on multi-task DRL, where a heuristic algorithm and a multi-task DRL framework were alternatively used to optimize device and UAV scheduling variables.
Finally, the theoretical analysis and performance results demonstrated that our solution has obvious advantages in convergence speed and average computation efficiency.

\section*{APPENDIX A}
{\appendices
	\section*{Proof of Theorem 1}
	In Problem P1, scheduling variables $ \alpha_m^t  $ and $ \beta_u^t $ are binary.
	The objective function is the ratio of long-term system computation bits to system energy consumption, expressed by equation (\ref{cet}).
	Constraints (\ref{p1}a) and (\ref{p1}b) are non-convex functions about device scheduling variable $ \alpha_m^t  $ and time allocation variable $ \tau^t $, and constraint (\ref{p1}c) is a non-convex function about UAV scheduling variable $ \beta_u^t $ and UAV location $ q_u^t $. 
	Meanwhile, the objective function is a non-convex function about $ \alpha_m^t  $, $ \tau^t $, $ \beta_u^t $ and $ q_u^t $.
	Thus, Problem P1 is a mixed integer non-convex fractional optimization problem.
	Even without considering time allocation variable $ \tau^t $ and UAV location $  q_u^t $, Problem P1 is still an integer linear fractional programming problem, which is also NP-hard~\cite{ref7}.
	As a result, Problem P1 is NP-hard.}

\section*{APPENDIX B}
{\appendices
	\section*{Proof of Theorem 2}
	Inequation (\ref{p1}a) can be transformed into the following inequation:
	\begin{align*}
		\tau^t \le \frac{E_{m}^{t-1,r}}{{(\alpha}_m^{t*}P_m+k_mf_m^3)T}.\tag{B.1}
	\end{align*}
	In addition, the detailed expression of inequation (\ref{p1}b) is inequation (B.2), and can be further converted to inequation (B.3) by mathematical deformation, as shown at the top of the next page.
	\begin{figure*}
		\begin{align*}
			k_mf_m^3T+(1-\tau^t-\alpha_m^{t*}+2\alpha_m^{t*}\tau^t)TP_m \le& E_{m}^{t-1,r}+\varepsilon(1-\alpha_m^{t*}) \widehat{h}_{m}^t\mathcal{P}_0\tau^tT+\varepsilon\alpha_m^{t*} \widehat{h}_{m}^t\mathcal{P}_0(1-\tau^t)T,\tag{B.2}\\
			\tau^t \ge&  {\frac{Tk_mf_m^3+T\left(1-\alpha_m^{t*}\right)P_m-E_{m}^{t-1,r}-\varepsilon T\alpha_m^{t*} \widehat{h}_{m}^t\mathcal{P}_0}{T\left(1-2\alpha_m^{t*}\right)\left(P_m+\varepsilon \widehat{h}_{m}^t\mathcal{P}_0\right)}}.\tag{B.3}
		\end{align*}
		\hrulefill
	\end{figure*}

Therefore, a further representation of Problem P2$' $ can be found in equation (B.4), which can be found at the top of the next page.
\begin{figure*}
	\begin{align*}
		&\mbox{P2}':\quad \underset { {\tau^ t}}{ \max}\quad \cfrac{\sum _{m =1}^{M} {L_m^t}'} {\sum _{m =1}^{M} {E_m^t}' + \sum _{u =1}^{U} E_u^t},\\
		&\quad\quad= \underset { {\tau^ t}}{ \max}  \frac{\tau^t\sum_{m=1}^{M}{(2\alpha_m^{t*}-1)}\mathbb{B}{log}_2\left(1+\frac{P_mh_{mu}^t}{\delta^2}\right)+\sum_{m=1}^{M}{\Big(\frac{f_m}{C_1}+(1-\alpha_m^{t*})\mathbb{B}{log}_2\left(1+\frac{P_mh_{mu}^t}{\delta^2}\right)\Big)}}{\tau^t\sum_{m=1}^{M}{P_m(2\alpha_m^{t*}-1)+\sum_{m=1}^{M}{{(k}_mf_m^3+P_m(1}-\alpha_m^{t*}))+\sum_{u=1}^{U}{\Big(\beta_u^tk_u{f_u}^3+ \zeta _{1}\| \mathbb{V}_u^t\|^{3}+\frac {\zeta _{2}}{\| \mathbb{V}_u^t\|}\Big)}}}, \tag{B.4}
		\\&\mbox{s.t.}\quad  \ \tau^t\le\frac{E_{m}^{t-1,r}}{{(\alpha}_m^{t*}P_m+k_mf_m^3)T},m\in \mathbb{M},\tag{B.4a}
		\\& \quad \quad \ \tau^t\geq\ \frac{Tk_mf_m^3+T\left(1-\alpha_m^{t*}\right)P_m-E_{m}^{t-1,r}-\varepsilon T\alpha_m^{t*} \widehat{h}_{m}^t\mathcal{P}_0}{T\left(1-2\alpha_m^{t*}\right)\left(P_m+\varepsilon \widehat{h}_{m}^t\mathcal{P}_0\right)},m\in \mathbb{M}.\tag{B.4b}
	\end{align*}
	\hrulefill
\end{figure*}

	By observing inequations (B.4a) and (B.4b), it can be noted that, to meet energy constraints of both type-1 and type-2 devices, time allocation variable $ \tau^t $ can be confined to the range as expressed in equation (B.5), which can be found at the top of the next page. 
	\begin{figure*}
		\begin{align*}\label{b5}
			&\tau^t\in(\underset { {m}}{ \max}\{{\frac{Tk_mf_m^3+T\left(1-\alpha_m^{t*}\right)P_m-E_{m}^{t-1,r}-\varepsilon T\alpha_m^{t*} \widehat{h}_{m}^t\mathcal{P}_0}{T\left(1-2\alpha_m^{t*}\right)\left(P_m+\varepsilon \widehat{h}_{m}^t\mathcal{P}_0\right)}}\},\underset { {m}}{ \min}\{{\frac{E_{m}^{t-1,r}}{{(\alpha}_m^{t*}P_m+k_mf_m^3)T}}\}).&\tag{B.5}
		\end{align*}
		\hrulefill
	\end{figure*} 

	Furthermore, we define  $ A=\sum_{m=1}^{M}{(2\alpha_m^{t\ast}-1)}\mathbb{B}{log}_2\left(1+{P_mh_{mu}^t}/{\delta^2}\right) $, $ B = \sum_{m=1}^{M}\Big({f_m}/{C_1}+(1-\alpha_m^{t\ast})\mathbb{B}{log}_2(1+{P_mh_{mu}^t}/{\delta^2})\Big) $, $ C=\sum_{m=1}^{M}{P_m(2\alpha_m^{t\ast}-1)} $ and $ D = \sum_{m=1}^{M}({k}_mf_m^3+P_m(1-\alpha_m^{t\ast}))+\sum_{u=1}^{U}\Big( \beta_u^tk_u{f_u}^3+ \zeta _{1}\| \mathbb{V}_u^t\|^{3}+ {\zeta _{2}}/{\| \mathbb{V}_u^t\|}\Big) $.
	Then, we can obtain function $ f(\tau^t) $ for Problem P2$' $ as equation (B.6), which can be found at the top of the next page. 
	\begin{figure*}
		\begin{align*}
			f(\tau^t)&= \frac{\tau^t\sum_{m=1}^{M}{(2\alpha_m^{t*}-1)}\mathbb{B}{log}_2\left(1+\frac{P_mh_{mu}^t}{\delta^2}\right)+\sum_{m=1}^{M}{\Big(\frac{f_m}{C_1}+(1-\alpha_m^{t*})\mathbb{B}{log}_2\left(1+\frac{P_mh_{mu}^t}{\delta^2}\right)\Big)}}{\tau^t\sum_{m=1}^{M}{P_m(2\alpha_m^{t*}-1)+\sum_{m=1}^{M}{{(k}_mf_m^3+P_m(1}-\alpha_m^{t*}))+\sum_{u=1}^{U}{\Big(\beta_u^tk_u{f_u}^3+ \zeta _{1}\| \mathbb{V}_u^t\|^{3}+\frac {\zeta _{2}}{\| \mathbb{V}_u^t\|}\Big)}}}\\
			&=\frac{{A\tau}^t+B}{C\tau^t+D}. \tag{B.6}
		\end{align*}
		\hrulefill
	\end{figure*} 
	
	\indent Further, the first-order derivative of $ f(\tau^t) $ can be computed by:
	\begin{align*}
	{f'\left(\tau^t\right)}= \frac{AD-BC}{{(C\tau^t+D)}^2}. \tag{B.7}
	\end{align*}

	 By observing equation (B.7), we can find that $ {f'\left(\tau^t\right)}  $ has no zero point, and thus $ f(\tau^t) $ is a monotonic function.
	Since the denominator of $ {f'\left(\tau^t\right)} $ is constantly bigger than 0, the polarity of $ {f'\left(\tau^t\right)} $ depends on the polarity of $ AD-BC $.
	When $ AD-BC >0$, $ f(\tau^t $) is a monotonically increasing function and the maximum value is obtained at the upper bound. 
	On the contrary, when $  AD-BC<0 $, $ f(\tau^t) $ is a monotonically decreasing function and the maximum value is obtained at the lower bound.
	The case of $ AD-BC=0$ does not exist, which is further explained in the proof of Corollary 1.
	Therefore, the optimal solution of Problem  P2$' $ is: 
	\begin{align*}\label{taut} 
		\tau^{t\ast}=
		\begin{cases}
			\underset { {m}}{ \min}\{\frac{E_{m}^{t-1,r}}{T{(\alpha}_m^{t\ast} P_m+k_mf_m^3)}\},\quad &AD-BC>0;\\
			\underset { {m}}{ \max}\{ \frac{k_mf_m^3+\left(1-\alpha_m^{t\ast}\right)P_m}{\left(1-2\alpha_m^{t\ast}\right)\left(P_m+\varepsilon \widehat{h}_{m}^t\mathcal{P}_0\right)} \\+\frac{-{E_{m}^{t-1,r}{-\varepsilon T\alpha}_m^{t\ast} \widehat{h}_{m}^t\mathcal{P}_0}}{T\left(1-2\alpha_m^{t\ast}\right)\left(P_m+\varepsilon \widehat{h}_{m}^t\mathcal{P}_0\right)} \},\quad &AD-BC<0.
		\end{cases}\tag{B.8}
\end{align*}}

\section*{APPENDIX C}
{\appendices
	\section*{Proof of Corollary 1}
	\indent According to Theorem 2, we know that the optimal value of $ \tau^t  $ is a segmental function concerning the polarity of $ AD-BC $, in which, variables $ A $, $ B $, $ C $, and $ D $ are defined in Theorem 2.
	Observing the expressions for variables $ A $ and $ C $, we find that $ \mathbb{B}{log}_2\left(1+{P_mh_{mu}^t}/{\delta^2}\right) $ and $ P_m $ are constantly bigger than 0, and variables $ A $ and $ C $ have common factorization $ (2\alpha_m^{t\ast}-1) $. 
	Therefore, variables $ A $ and $ C $ have the same polarity. 
	
By multiplying $ B $ by time interval $ T $, the following equation can be obtained by:
	\begin{align*}
		&B\times T=\sum_{m=1}^{M}\Bigg({\frac {f_mT}{C_1}+(1-\alpha_m^{t\ast})T\mathbb{B}{log}_2\left(1+\frac{P_mh_{mu}^t}{\delta^2}\right)}\Bigg). &\tag{C.1}
	\end{align*}
	The physical meaning of equation (C.1) is the number of local computation bits for all devices plus the number of offloaded data bits for type-2 devices. 
	Therefore, variable $ B $ is constantly bigger than 0.
	
By multiplying $ D $ by time interval $ T $, the following equation can be obtained by:
	\begin{align*}
		D\times T&= \sum_{m=1}^{M}{{T(k}_mf_m^3+P_m(1}-\alpha_m^{t\ast}))
		\\&+\sum_{u=1}^{U}T{\Bigg(\beta_u^tk_u{f_u}^3+ \zeta _{1}\| \mathbb{V}_u^t\|^{3}+\frac {\zeta _{2}}{\| \mathbb{V}_u^t\|}\Bigg)}\\
		&=\sum_{m=1}^{M}(E_{m}^{t,l}+TP_m(1-\alpha_m^{t\ast}))+\sum_{u=1}^{U}\mathbb{E}_u^t,\tag{C.2}
	\end{align*}
	where $ E_{m}^{t,l} $ and $ E_u^t $ denote the local energy consumption of device $ m $ and the total energy consumption of UAV $ u $ in time slot $ t$, respectively.
	Furthermore, expression $ T P_m\left(1-\alpha_m^{t\ast}\right)  $ is denoted as the transmission energy consumption of type-2 device $ m $ in time slot $ t $.
	Therefore, variable $ D $ is constantly bigger than 0. 
	
By multiplying variable $ C$ by time interval $ T $, the following equation can be obtained by:
	\begin{align*}
		C\times T&= \sum_{m=1}^{M}T{P_m(2\alpha_m^{t\ast}-1)}
		\\&=\sum_{m=1}^{M}(\alpha_m^{t\ast}T{P_m} -(1-\alpha_m^{t\ast})T{P_m}).\tag{C.3}
	\end{align*}
	The physical meaning of variable $ C $ times $ T $ is the sum of the transmission energy consumption of all type-1 devices minus the sum of the transmission energy consumption of all type-2 devices.
	The flight energy consumption of UAVs is known to be one or two orders of magnitude bigger than the transmission energy consumption of the mobile device~\cite{ref35}. 
	Furthermore, $ DT-CT $ is constantly bigger than 0.
	Therefore, variable $ D $ is constantly bigger than variable $ C $.
	By observing the expressions of variables $ A $ and $ B $, we find that they belong to the same order of magnitude.
	
	As a result, the polarity of $ AD-BC $ depends only on the polarity of variables $ A $ and $ C $.
	By the previous derivation, we can find that the polarity of  variable $ C $ is consistent with variable $ A $. 
	Therefore, Corollary 1 is proved.}

\section*{APPENDIX D}
{\appendices
	\section*{Proof of Theorem 3}
	Based on Theorem 2 and Corollary 1, we can derive the following equations:
	\begin{align*}\label{taut} 
		\tau^{t\ast}=
		\begin{cases}
			\underset { {m}}{ \min}\{\frac{E_{m}^{t-1,r}}{T{(\alpha}_m^{t\ast} P_m+k_mf_m^3)}\},\quad &AD-BC>0;\\
			\underset { {m}}{ \max}\{ \frac{k_mf_m^3+\left(1-\alpha_m^{t\ast}\right)P_m}{\left(1-2\alpha_m^{t\ast}\right)\left(P_m+\varepsilon \widehat{h}_{m}^t\mathcal{P}_0\right)} \\+\frac{-{E_{m}^{t-1,r}{-\varepsilon T\alpha}_m^{t\ast} \widehat{h}_{m}^t\mathcal{P}_0}}{T\left(1-2\alpha_m^{t\ast}\right)\left(P_m+\varepsilon \widehat{h}_{m}^t\mathcal{P}_0\right)} \},\quad &AD-BC<0.
		\end{cases}\tag{D.1}
	\end{align*}
	It is known that the expression for variable $ A $  is denoted by:
	\begin{align*}
		A&=\sum_{m=1}^{M}{(2\alpha_m^{t\ast}-1)}\mathbb{B}{log}_2\left(1+\frac{P_mh_{mu}^t}{\delta^2}\right)\\&=\sum_{m=1}^{M}	\Bigg(\alpha_m^{t\ast}\mathbb{B}{log}_2\left(1+\frac{P_mh_{mu}^t}{\delta^2}\right)\\&- (1-\alpha_m^{t\ast})\mathbb{B}{log}_2\left(1+\frac{P_mh_{mu}^t}{\delta^2}\right) \Bigg). \tag{D.2}
	\end{align*}
	
	 It is known that expression $ \mathbb{B}{log}_2\left(1+{P_mh_{mu}^t}/{\delta^2}\right) $ is denoted as the maximum channel capacity between UAV $ u $ and mobile device $ m $ in time slot $ t $, and $ \alpha_m^{t*} $ is equal to $1$ or $0$ indicating that device $ m $ is classified as the type-1 device or the type-2 device, respectively.
	Then, variable $ A $ defined in equation (D.2) can be regarded by the sum of channel capacity of all type-1 devices in time slot $ t $ minus that of all type-2 devices.
	When the sum of channel capacity of all type-1 devices equals to that of all type-2 devices, variable $ A $ equals to $0$.
	In that case, $  \tau^t $ does not affect the value of $ f(\tau^t) $ defined in equation (B.6).
	However, since we consider the scenario where devices move randomly within the region, it is almost impossible that the sum of channel capacity of all type-1 devices happens to be equal to that of all type-2 devices.
	Therefore, we do not consider the case when variable $ A $ equals to $0$.

	Finally, when variable $ A $ is bigger than 0, it means that the sum of channel capacity of type-1 devices is bigger than that of all type-2 devices, i.e., $ AD-BC>0 $, and time allocation variable $  \tau^{t*} $ can reach the upper bound of the value domain.
	Conversely, when variable $ A $ is less than 0, it means that the sum of channel capacity of type-1 devices is less than that of all type-2 devices, i.e., $ AD-BC<0 $, and time allocation variable $ \tau^{t*}  $ can reach the lower bound of the value domain.}

\section*{APPENDIX E}
{\appendices
	\section*{PROOF of Theorem 4}
	Similar to~\cite{ref15}, the complexity of the learning agent is mainly related to the configuration of the neural network.
Let $ \imath_{u,i} $ and $ \imath_{0,i} $ represent the number of neurons at layer $ i $ of UAV-specific policy network $ u $ and the shared policy network, respectively.
	Thus, the total complexity of the shared policy network and UAV-specific policy networks is $  \mathcal {O}\Big( \sum \limits _{u=1}^{U} \sum \limits _{i=1}^{\Re -1} \imath_{u,i}\times \imath _{u,i+1} + \sum \limits _{j=1}^{\Im -1} \imath _{0,i}\times \imath _{0,i+1} \Big) $, where $ \Re $ and $ \Im $ are the number of fully connected layers of each UAV-specific policy network and the shared policy network.
	Since UAV-specific policy networks and the shared policy network are alternately optimized and $ B $ experience is extracted from replay buffer, the complexity of the training process of the designed multi-task DRL-based algorithm is $ \mathcal {O}\Big(\vphantom{\sum \limits _{i=1}^{\Re -1}} L \times E \times B \times \Big( \sum \limits _{u=1}^{U} \sum \limits _{i=1}^{\Re -1} \imath_{u,i}\times \imath _{u,i+1} + \sum \limits _{j=1}^{\Im -1} \imath _{0,i}\times \imath _{0,i+1} \Big) \Big) $.
	Therefore, Theorem 4 is proved.
}

\section*{APPENDIX F}
{\appendices
	\section*{PROOF Theorem 5}
	
	The complexity of Algorithm 2 comes from two aspects.
	The first aspect is from computation scheduling and time allocation for mobile devices.
	The second aspect is from the execution of the learning algorithm to obtain UAV scheduling and trajectory design decision.
	Let $ N $ denote the number of time slots required for the execution of MURAL algorithm.
	Meanwhile, $ L_t,\;t\in(0,N) $, denotes the number of iterations required to obtain the optimal scheduling results for both UAVs and mobile devices in time slot $ t $.
	The complexity of the designed heuristic algorithm to realize computation scheduling and time allocation for mobile devices is $ \mathcal{O}(M) $, where $ M $ is the number of mobile devices.
	In addition, charging scheduling and trajectory design decision for UAVs can be obtained from trained UAV-specific networks with the complexity of $ \mathcal{O}\Big(\sum \limits _{u=1}^{U} \sum \limits _{i=1}^{\Re -1} \imath_{u,i}\times \imath _{u,i+1}\Big) $.
	Therefore, the complexity of Algorithm 2 is $ \mathcal{O}\Big(\sum_{t=0}^{N} L_t\Big(\sum \limits _{u=1}^{U} \sum \limits _{i=1}^{\Re -1} \imath_{u,i}\times \imath _{u,i+1}+M\Big)\Big)$.

\bibliographystyle{IEEEtran}
\bibliography{reference}

\begin{thebibliography}{10}
\providecommand{\url}[1]{#1}
\csname url@samestyle\endcsname
\providecommand{\newblock}{\relax}
\providecommand{\bibinfo}[2]{#2}
\providecommand{\BIBentrySTDinterwordspacing}{\spaceskip=0pt\relax}
\providecommand{\BIBentryALTinterwordstretchfactor}{4}
\providecommand{\BIBentryALTinterwordspacing}{\spaceskip=\fontdimen2\font plus
\BIBentryALTinterwordstretchfactor\fontdimen3\font minus
  \fontdimen4\font\relax}
\providecommand{\BIBforeignlanguage}[2]{{%
\expandafter\ifx\csname l@#1\endcsname\relax
\typeout{** WARNING: IEEEtran.bst: No hyphenation pattern has been}%
\typeout{** loaded for the language `#1'. Using the pattern for}%
\typeout{** the default language instead.}%
\else
\language=\csname l@#1\endcsname
\fi
#2}}
\providecommand{\BIBdecl}{\relax}
\BIBdecl

\bibitem{ref19}
M.~Xu, W.~C. Ng, W.~Y.~B. Lim, J.~Kang, Z.~Xiong, D.~Niyato, Q.~Yang, X.~Shen,
  and C.~Miao, ``A full dive into realizing the edge-enabled metaverse:
  Visions, enabling technologies, and challenges,'' \emph{IEEE Communications
  Surveys \& Tutorials}, vol.~25, no.~1, pp. 656--700, Nov. 2023.

\bibitem{ref18}
R.~Hare and Y.~Tang, ``Hierarchical deep reinforcement learning with experience
  sharing for metaverse in education,'' \emph{IEEE Transactions on Systems,
  Man, and Cybernetics: Systems}, vol.~53, no.~4, pp. 2047--2055, Apr. 2023.

\bibitem{8967118}
Y.~Zhan, S.~Guo, P.~Li, and J.~Zhang, ``A deep reinforcement learning based
  offloading game in edge computing,'' \emph{IEEE Transactions on Computers},
  vol.~69, no.~6, pp. 883--893, Jun.2020.

\bibitem{9210202}
S.~Pan, P.~Li, C.~Yi, D.~Zeng, Y.-C. Liang, and G.~Hu, ``Edge intelligence
  empowered urban traffic monitoring: A network tomography perspective,''
  \emph{IEEE Transactions on Intelligent Transportation Systems}, vol.~22,
  no.~4, pp. 2198--2211, Apr. 2021.

\bibitem{8887204}
Q.~Xu, Z.~Su, K.~Zhang, and P.~Li, ``Intelligent cache pollution attacks
  detection for edge computing enabled mobile social networks,'' \emph{IEEE
  Transactions on Emerging Topics in Computational Intelligence}, vol.~4,
  no.~3, pp. 241--252, Jun. 2020.

\bibitem{ref21}
X.~Wang, J.~Li, Z.~Ning, Q.~Song, L.~Guo, S.~Guo, and M.~S. Obaidat, ``Wireless
  powered mobile edge computing networks: A survey,'' \emph{ACM Comput. Surv.},
  vol.~55, no. 13s, Jul. 2023.

\bibitem{ref23}
Q.~Liu, L.~Shi, L.~Sun, J.~Li, M.~Ding, and F.~Shu, ``Path planning for
  {UAV}-mounted mobile edge computing with deep reinforcement learning,''
  \emph{IEEE Transactions on Vehicular Technology}, vol.~69, no.~5, pp.
  5723--5728, May. 2020.

\bibitem{10239498}
Z.~Ning, Y.~Yang, X.~Wang, Q.~Song, L.~Guo, and A.~Jamalipour, ``Multi-agent
  deep reinforcement learning based {UAV} trajectory optimization for
  differentiated services,'' \emph{IEEE Transactions on Mobile Computing}, pp.
  1--17, Sep. 2023, doi: {10.1109/TMC.2023.3312276}.

\bibitem{ref35}
Z.~Yang, S.~Bi, and Y.-J.~A. Zhang, ``Stable online offloading and trajectory
  control for {UAV}-enabled {MEC} with {EH} devices,'' in \emph{Proc. IEEE
  GLOBECOM}, Dec. 2021, pp. 01--07.

\bibitem{ref36}
W.~Feng, J.~Tang, N.~Zhao, X.~Zhang, X.~Wang, K.-K. Wong, and J.~A. Chambers,
  ``Hybrid beamforming design and resource allocation for {UAV}-aided
  wireless-powered mobile edge computing networks with {NOMA},'' \emph{IEEE
  Journal on Selected Areas in Communications}, vol.~39, no.~11, pp.
  3271--3286, Nov. 2021.

\bibitem{ref37}
W.~Liu, S.~Zhang, and N.~Ansari, ``Joint laser charging and {DBS} placement for
  drone-assisted edge computing,'' \emph{IEEE Transactions on Vehicular
  Technology}, vol.~71, no.~1, pp. 780--789, Jan. 2022.

\bibitem{ref6}
X.~Hu, K.-K. Wong, and Y.~Zhang, ``Wireless-powered edge computing with
  cooperative {UAV}: Task, time scheduling and trajectory design,'' \emph{IEEE
  Transactions on Wireless Communications}, vol.~19, no.~12, pp. 8083--8098,
  Dec. 2020.

\bibitem{ref39}
Z.~Xu, K.~Wu, Z.~Che, J.~Tang, and J.~Ye, ``Knowledge transfer in multi-task
  deep reinforcement learning for continuous control,'' in \emph{Advances in
  Neural Information Processing Systems}, vol.~33, 2020, pp. 15\,146--15\,155.

\bibitem{ref9}
Y.~Teh, V.~Bapst, W.~M. Czarnecki, J.~Quan, J.~Kirkpatrick, R.~Hadsell,
  N.~Heess, and R.~Pascanu, ``Distral: Robust multitask reinforcement
  learning,'' in \emph{Advances in Neural Information Processing Systems},
  vol.~30, 2017, pp. 4499--4509.

\bibitem{ref40}
M.~Hessel, H.~Soyer, L.~Espeholt, W.~Czarnecki, S.~Schmitt, and H.~van Hasselt,
  ``Multi-task deep reinforcement learning with {PopArt},'' in
  \emph{Proceedings of the AAAI Conference on Artificial Intelligence},
  vol.~33, no.~01, 2019, pp. 3796--3803.

\bibitem{2018Belletti}
F.~Belletti, D.~Haziza, G.~Gomes, and A.~M. Bayen, ``Expert level control of
  ramp metering based on multi-task deep reinforcement learning,'' \emph{IEEE
  Transactions on Intelligent Transportation Systems}, vol.~19, no.~4, pp.
  1198--1207, Apr. 2018.

\bibitem{2023Zhang}
Z.~Zhang, F.~Zhuang, H.~Zhu, C.~Li, H.~Xiong, Q.~He, and Y.~Xu, ``Towards
  robust knowledge graph embedding via multi-task reinforcement learning,''
  \emph{IEEE Transactions on Knowledge and Data Engineering}, vol.~35, no.~4,
  pp. 4321--4334, Apr. 2023.

\bibitem{ref31}
Q.~Qi, L.~Zhang, J.~Wang, H.~Sun, Z.~Zhuang, J.~Liao, and F.~R. Yu, ``Scalable
  parallel task scheduling for autonomous driving using multi-task deep
  reinforcement learning,'' \emph{IEEE Transactions on Vehicular Technology},
  vol.~69, no.~11, pp. 13\,861--13\,874, Nov. 2020.

\bibitem{ref28}
T.~Dong, Z.~Zhuang, Q.~Qi, J.~Wang, H.~Sun, F.~R. Yu, T.~Sun, C.~Zhou, and
  J.~Liao, ``Intelligent joint network slicing and routing via {GCN}-powered
  multi-task deep reinforcement learning,'' \emph{IEEE Transactions on
  Cognitive Communications and Networking}, vol.~8, no.~2, pp. 1269--1286, Jun.
  2022.

\bibitem{ref29}
J.~Chen, S.~Chen, Q.~Wang, B.~Cao, G.~Feng, and J.~Hu, ``{IRAF}: A deep
  reinforcement learning approach for collaborative mobile edge computing {IoT}
  networks,'' \emph{IEEE Internet of Things Journal}, vol.~6, no.~4, pp.
  7011--7024, Apr. 2019.

\bibitem{ref2}
L.~Huang, S.~Bi, and Y.-J.~A. Zhang, ``Deep reinforcement learning for online
  computation offloading in wireless powered mobile-edge computing networks,''
  \emph{IEEE Transactions on Mobile Computing}, vol.~19, no.~11, pp.
  2581--2593, Nov. 2020.

\bibitem{ref3}
F.~Zhou, Y.~Wu, R.~Q. Hu, and Y.~Qian, ``Computation rate maximization in
  {UAV}-enabled wireless-powered mobile-edge computing systems,'' \emph{IEEE
  Journal on Selected Areas in Communications}, vol.~36, no.~9, pp. 1927--1941,
  Sep. 2018.

\bibitem{2021ning}
Z.~Ning, Y.~Yang, X.~Wang, L.~Guo, X.~Gao, S.~Guo, and G.~Wang, ``Dynamic
  computation offloading and server deployment for {UAV}-enabled multi-access
  edge computing,'' \emph{IEEE Transactions on Mobile Computing}, vol.~22,
  no.~5, pp. 2628--2644, May. 2023.

\bibitem{2021wang}
X.~Wang, Z.~Ning, S.~Guo, M.~Wen, L.~Guo, and H.~V. Poor, ``Dynamic {UAV}
  deployment for differentiated services: A multi-agent imitation learning
  based approach,'' \emph{IEEE Transactions on Mobile Computing}, vol.~22,
  no.~4, pp. 2131--2146, Apr. 2023.

\bibitem{ref1}
X.~Wang, Z.~Ning, L.~Guo, S.~Guo, X.~Gao, and G.~Wang, ``Online learning for
  distributed computation offloading in wireless powered mobile edge computing
  networks,'' \emph{IEEE Transactions on Parallel and Distributed Systems},
  vol.~33, no.~8, pp. 1841--1855, Aug. 2022.

\bibitem{9453824}
Z.~Ning, P.~Dong, M.~Wen, X.~Wang, L.~Guo, R.~Y.~K. Kwok, and H.~V. Poor,
  ``{5G}-enabled {UAV}-to-community offloading: Joint trajectory design and
  task scheduling,'' \emph{IEEE Journal on Selected Areas in Communications},
  vol.~39, no.~11, pp. 3306--3320, Jun. 2021.

\bibitem{ref7}
W.~Ejaz, M.~Naeem, and S.~Zeadally, ``On-demand sensing and wireless power
  transfer for self-sustainable industrial {I}nternet of things networks,''
  \emph{IEEE Transactions on Industrial Informatics}, vol.~17, no.~10, pp.
  7075--7084, Oct. 2021.

\bibitem{ref8}
A.~Tomar, L.~Muduli, and P.~K. Jana, ``A fuzzy logic-based on-demand charging
  algorithm for wireless rechargeable sensor networks with multiple chargers,''
  \emph{IEEE Transactions on Mobile Computing}, vol.~20, no.~9, pp. 2715--2727,
  Sep. 2021.

\bibitem{ref11}
O.~Nachum, M.~Norouzi, K.~Xu, and D.~Schuurmans, ``Bridging the gap between
  value and policy based reinforcement learning,'' in \emph{Advances in Neural
  Information Processing Systems}, vol.~30, 2017, pp. 2772--2782.

\bibitem{ref15}
O.~S. Oubbati, M.~Atiquzzaman, H.~Lim, A.~Rachedi, and A.~Lakas,
  ``Synchronizing {UAV} teams for timely data collection and energy transfer by
  deep reinforcement learning,'' \emph{IEEE Transactions on Vehicular
  Technology}, vol.~71, no.~6, pp. 6682--6697, Jun. 2022.

\bibitem{ref16}
S.~Zhang, H.~Gu, K.~Chi, L.~Huang, K.~Yu, and S.~Mumtaz, ``{DRL}-based partial
  offloading for maximizing sum computation rate of wireless powered mobile
  edge computing network,'' \emph{IEEE Transactions on Wireless
  Communications}, vol.~21, no.~12, pp. 10\,934--10\,948, Dec. 2022.

\end{thebibliography}

\begin{IEEEbiography}[{\includegraphics[width=1in,height=1.25in,clip,keepaspectratio]{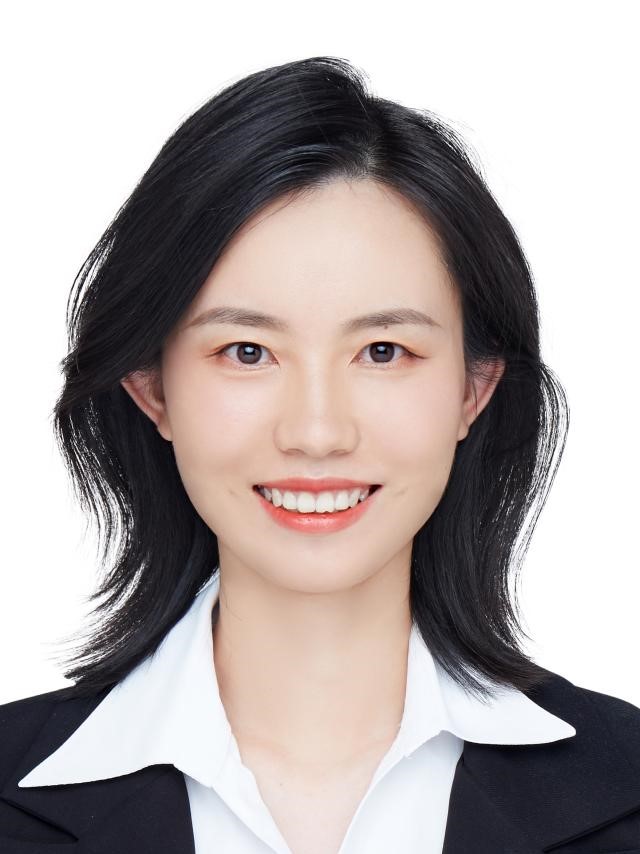}}]{Xiaojie Wang}
	(M'19-SM'23) received the PhD degree from Dalian University of Technology, Dalian, China, in 2019. After that, she was a postdoctor in the Hong Kong Polytechnic University. Currently, she is a full professor with the School of Communication and Information Engineering, the Chongqing University of Posts and Telecommunications, Chongqing, China. Her research interests are wireless networks, mobile edge computing and machine learning. She has published over 60 scientific papers in international journals and conferences, such as IEEE TMC, IEEE JSAC, IEEE TPDS and IEEE COMST.
\end{IEEEbiography}
\vspace{-20 pt} 
\begin{IEEEbiography}[{\includegraphics[width=1in,height=1.25in,clip,keepaspectratio]{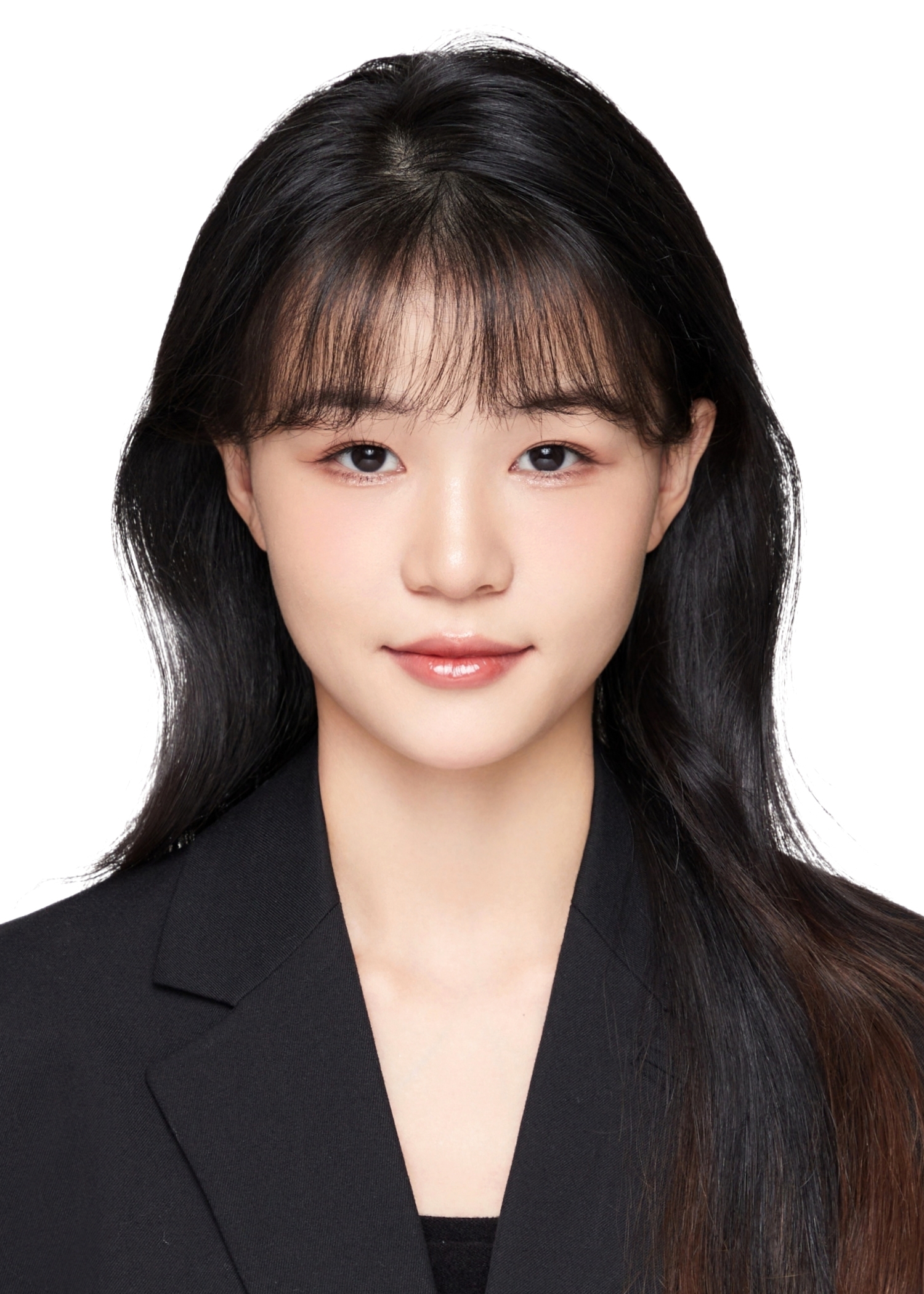}}]{Jiameng Li}
	received the B.E. degree in electronic information science and technology from the Chongqing Normal University, Chongqing, China in 2021.
	She is currently working toward the M.E. degree with the School of Communication and Information Engineering, Chongqing University of Posts and Telecommunications, Chongqing, China. 
	Her research interests include mobile edge computing, unmanned aerial vehicle and wireless power transfer.
\end{IEEEbiography}
\vspace{-20 pt} 
\begin{IEEEbiography}[{\includegraphics[width=1in,height=1.25in,clip,keepaspectratio]{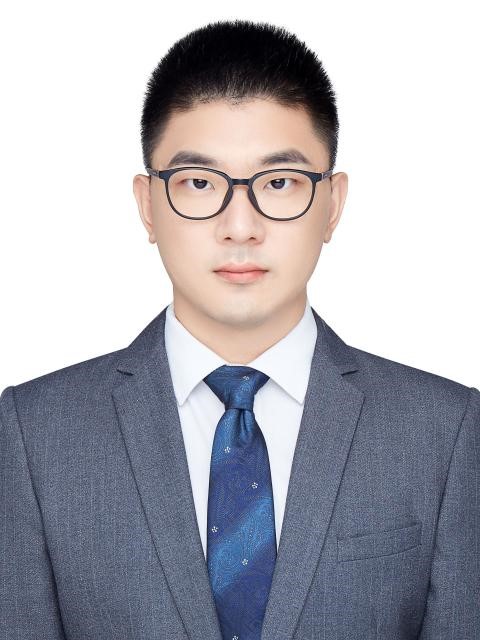}}]{Zhaolong Ning}
	(M'14-SM'18) received the Ph.D. degree from Northeastern University, China in 2014. He was a Research Fellow at Kyushu University from 2013 to 2014, Japan. Currently, he is a full professor with the School of Communication and Information Engineering, the Chongqing University of Posts and Telecommunications, Chongqing, China. His research interests include mobile edge computing, 6G networks, machine learning, and resource management. He has published over 150 scientific papers in international journals and conferences. Dr. Ning serves as an associate editor or guest editor of several journals, such as IEEE Transactions on Vehicular Technology, IEEE Transactions on Industrial Informatics, IEEE Transactions on Social Computational Systems, IEEE Internet of Things Journal and so on. He is a Highly Cited Researcher (Web of Science) since 2020.
\end{IEEEbiography}
\vspace{-20 pt} 
\begin{IEEEbiography}[{\includegraphics[width=1in,height=1.25in,clip,keepaspectratio]{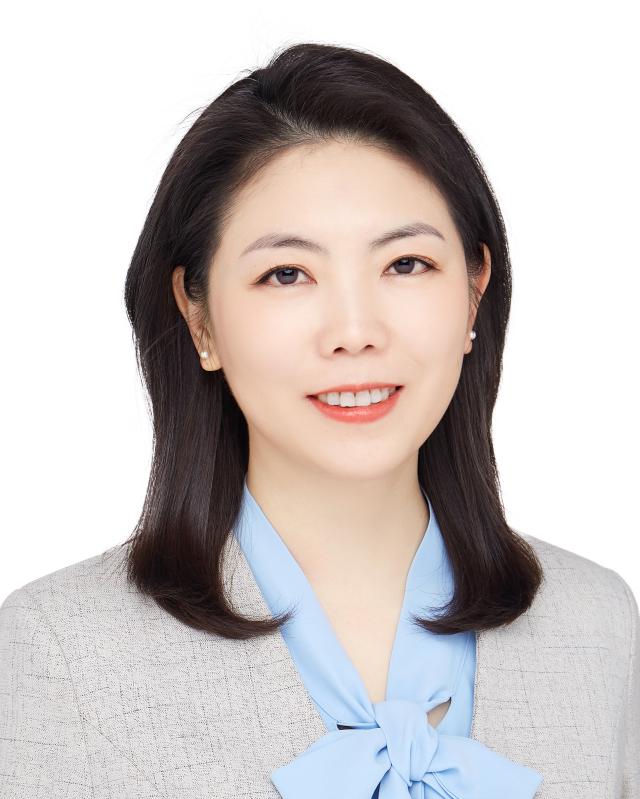}}]{Qingyang Song}
	(S'05-M'07-SM'14) received the Ph.D. degree in telecommunications engineering from the University
	of Sydney, Australia, in 2007. From 2007 to 2018, she was with Northeastern University, China. She joined Chongqing
	University of Posts and Telecommunications in 2018, where she is currently a Professor. She has authored more
	than 100 papers in major journals and international conferences. Her current research interests include cooperative
	resource management, edge computing, connected and autonomous vehicles systems. She serves on the editorial boards of two journals, including Area Editor for IEEE Transactions on Vehicular Technology and Academic Editor for Digital Communications and Networks.
\end{IEEEbiography}
\vspace{-20 pt} 
\begin{IEEEbiography}[{\includegraphics[width=1in,height=1.25in,clip,keepaspectratio]{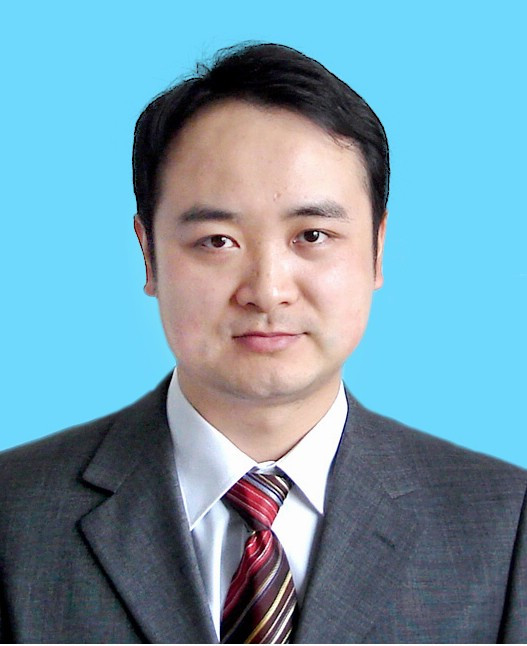}}]{Lei Guo}
 received the Ph.D. degree from the University of Electronic Science and Technology of China, Chengdu, China, in 2006. He is currently a full professor with Chongqing University of Posts and Telecommunications, Chongqing, China. He has authored or coauthored more than 200 technical papers in international journals and conferences. He is an editor for several international journals. His current research interests include communication networks, optical communications, and wireless communications.	
\end{IEEEbiography}
\vspace{-20 pt} 
\begin{IEEEbiography}[{\includegraphics[width=1in,height=1.25in,clip,keepaspectratio]{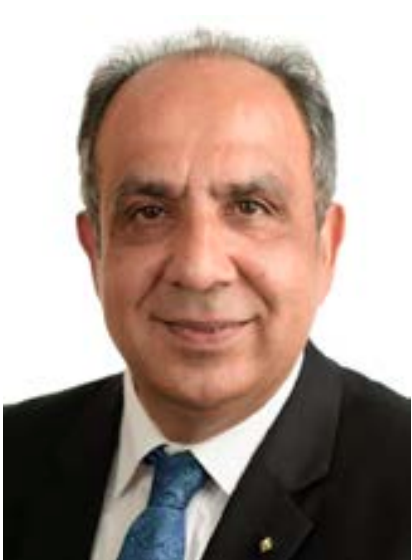}}]{Abbas Jamalipour}
	(S’86-M’91-SM’00-F’07) received the Ph.D. degree in Electrical Engineering from Nagoya University, Nagoya, Japan in
	1996. He holds the positions of Professor of
	Ubiquitous Mobile Networking with the University of Sydney and since January 2022, the
	Editor-in-Chief of the IEEE Transactions on Vehicular Technology. He has authored nine technical books, eleven book chapters, over 550 technical papers, and five patents, all in the area of
	wireless communications and networking. Prof.
	Jamalipour is a recipient of the number of prestigious awards, such as
	the 2019 IEEE ComSoc Distinguished Technical Achievement Award in
	Green Communications, the 2016 IEEE ComSoc Distinguished Technical Achievement Award in Communications Switching and Routing, the
	2010 IEEE ComSoc Harold Sobol Award, the 2006 IEEE ComSoc Best
	Tutorial Paper Award, as well as over 15 Best Paper Awards. He was
	the President of the IEEE Vehicular Technology Society (2020-2021).
	Previously, he held the positions of the Executive Vice-President and the
	Editor-in-Chief of VTS Mobile World and has been an elected member
	of the Board of Governors of the IEEE Vehicular Technology Society
	since 2014. He was the Editor-in-Chief IEEE WIRELESS COMMUNICATIONS, the Vice President-Conferences, and a member of Board of
	Governors of the IEEE Communications Society. He sits on the Editorial
	Board of the IEEE ACCESS and several other journals and is a member
	of Advisory Board of IEEE Internet of Things Journal. He has been
	the General Chair or Technical Program Chair for several prestigious
	conferences, including IEEE ICC, GLOBECOM, WCNC, and PIMRC. He
	is a Fellow of the Institute of Electrical and Electronics Engineers (IEEE),
	the Institute of Electrical, Information, and Communication Engineers
	(IEICE), and the Institution of Engineers Australia, an ACM Professional
	Member, and an IEEE Distinguished Speaker.
\end{IEEEbiography}

\newpage

\vspace{11pt}

%

\vfill

\end{document}